\documentclass[11pt,titlepage]{article} %for JASA 12pt
\usepackage[dvips]{graphicx}
\usepackage{hyperref}

\usepackage{layout}

\newtheorem{Theorem}{Theorem}[section]

\newcommand{\bfbeta}{\mbox{\boldmath $\beta$}}
\newcommand{\bfalpha}{\mbox{\boldmath $\alpha$}}

\newcommand{\bftheta}{\mbox{\boldmath $\theta$}}
\newcommand{\bfrho}{\mbox{\boldmath $\rho$}}
\newcommand{\bfz}{\mbox{\boldmath $z$}}
\newcommand{\bfX}{\mbox{\boldmath $X$}}
\newcommand{\bft}{\mbox{\boldmath $t$}}
\newcommand{\bfh}{\mbox{\boldmath $h$}}

\newcommand*{\ackname}{\mbox{\bf Acknowledgements}:}

\usepackage{setspace}
\begin{document}

%Your username: bkedem
%Your new password: 1SubYqqO

\begin{titlepage}

\begin{center}
\mbox{\Large{\bf Repeated out of Sample Fusion in the Estimation}}\\
\mbox{\Large{\bf of  Small Tail Probabilities}}\\
\smallskip
\smallskip
\smallskip
\smallskip
Benjamin Kedem, Lemeng Pan, Paul Smith,  and Chen Wang\\
\smallskip
\smallskip
\smallskip
\smallskip
Department of Mathematics\\ 
%and Institute for Systems Research, 
University of Maryland, College Park\\
\smallskip
\smallskip
\smallskip
\smallskip
%February 2016
%March 2017
%March 2018
%May 2018
%July 2018
%August 2018
%Oct 2018
%November 2018
%April 2019
May 2019
\end{center}

\end{titlepage}

\begin{abstract}

%In pursuit of a small tail probability $p$,
%it is shown how to construct by repeated fusion of a given reference sample  
%with computer generated samples
%the distribution function of an upper bound for $p$. 
%interval estimates for $p$. 
%are obtained
%by tuning the distribution function.
%A comparison is made with an extreme value theory method for tail estimation
%using data from distributions with different types of tail behavior.

Often, it is required to estimate the probability that a quantity such as
toxicity level, plutonium, temperature, rainfall, damage, wind
speed, wave size, earthquake magnitude, risk, etc., exceeds an unsafe high threshold. 
The probability in
question is then very small. To estimate such a probability, information is needed
about large values of the quantity of interest. However, in many cases, the data
only contain values below or even far below the designated threshold, let alone exceedingly
large values. 
%which ostensibly renders the problem insolvable. 
It is shown that
by repeated fusion of the data with externally generated random data, more
information about small tail probabilities is obtained with the aid of certain new
statistical functions. This provides relatively short, yet reliable interval estimates based on
moderately large samples. A comparison of the approach with a method from
extreme values theory (peaks over threshold, or POT), using both artificial and
real data, points to the merit of repeated out of sample fusion.
\end{abstract}

\section{Introduction}

%Method can fail if max(x_0) is too small. We say this in the discussion.

%%Feldman: One gets two random numbers. We see only one number and must guess the 2nd. 
%%Generating Random numbers help!!!

This paper addresses the following basic problem.
Consider a moderately large random 
sample $\bfX_0=(X_1,...,X_{n_{0}})$ where all the observations are much smaller than 
a high threshold $T$, that is $\max(\bfX_0) < < T$. Based on the sample we
wish to estimate the probability $p$ of exceeding $T$ without knowing the underlying distribution.
However, 
as is, the sample may not contain sufficient amount of information to tackle the problem.
To gain more information, the problem is approached by combining or fusing the sample repeatedly with  externally generated computer data.

The problem is inspired by real world situations where 
the measurements $X_i$ are all below $T$. As an example, consider a sample of
rogue ocean waves none of which exceeds $T=150$ feet in height, and yet we wish to estimate the 
small chance of
exceeding $T$ from a moderately large sample $\bfX_0=(X_1,...,X_{n_{0}})$, referred to as a 
{\em reference} sample. Similar problems pertain to insurance claims, food safety, 
and environmental risks such as radiation levels.

How is this done? 
This article advances a statistical notion where
a small tail probability is identified with a {\em point} on a certain monotone curve
called {\em B-curve}, obtained by repeated fusion of real and artificial data.
The point on the curve is approached by an iterative procedure, against the backdrop of 
numerous fusions of
real and computer generated data, leading to a certain random ``fixed point"
for a lack of a better term.

Let  
$\bfX_i$ denote the $i$th computer generated sample of size $n_1=n_0$. 
Then the {\em fused} or {\em combined} samples are the {\em augmentations}
\begin{eqnarray}
(\bfX_0,\bfX_1),(\bfX_0,\bfX_2),(\bfX_0,\bfX_3)...
\label{RepeatedFusion}
\end{eqnarray}
where $\bfX_0$ is a real reference sample and the $\bfX_i$ are different independent computer generated samples.
The number of fusions can be as large as we wish.
For example 10,000 or 100,000 or 1,000,000 or more fusions.

{\em In that way, by repeated fusion, we can extract information about tail probabilities possibly not available in the original reference sample by itself.}
As we shall see,
in many cases this brings about surprisingly precise estimates for small tail 
probabilities, using moderately large samples (e.g. 100 or 200 data points), 
as described and illustrated in Section \ref{Capturing a Point on the B-Curve}.

Fusing a given sample repeatedly with computer generated data is referred to as
{\em repeated out of sample fusion (ROSF)} (Kedem et al. 2016). Unlike the bootstrap, additional information 
is sought repeatedly from outside the sample.
Related ideas concerning a single fusion are studied in
Fithian and Wager (2015), Fokianos and Qin (2008), Katzoff et al. (2014), 
 and Zhou (2013).

\subsection{An Iterative Procedure and its Estimates}

As said,
the large number of fusions results in a {\em B-curve}
defined in Section \ref{ROSF and the B-Curve}. The B-curve 
is monotonically increasing and it contains a {\em point}
whose ordinate is very close to $p$ with a high probability. As the number of fusions
increases the ordinate of that point essentially coincides with $p$. The goal is to ``capture"
that point by an iterative algorithm.

%Since the B-curve is monotone, 
%Estimating $p$ is equivalent to 
%``capturing" 
%(approximately)
%the said point on the B-curve, and this paper provides an iterative algorithm for doing so. 
%the point of interest on the B-curve. 
The consequential estimates as well as interval estimates of $p$ 
are quite precise.
A comparison with peaks-over-threshold (POT) from extreme value theory 
(Beirlant et al. 2004, Ferreira and DeHaan 2015)
indicates that
ROSF can bring about a substantial gain
in reliability as well as in precision across a 
fairly wide range of tail behavior, given moderately large samples $\bfX_0$. 
%See Kedem et al. (2018).

The question then is how to tie or connect
the real data and the generated random data 
to obtain useful reliable estimates
for small tail probabilities. 
Connecting or fusing the real and artificial data can be approached  
by means of their respective probability distributions 
under the so called {\em density ratio model} framework,
discussed briefly in Section \ref{Getting Upper Bounds by Data Fusion}
and in the Appendix.

Thus, the paper describes ROSF and a related iterative method (IM) in the estimation of small
tail probabilities, against the backdrop of the density ratio model, by ``capturing" a point on 
the B-curve.

\subsection{A Typical ``Down-Up" Example}

We illuminate our method upfront by a typical example, 
postponing technical details to later sections.
Let $\bfX_0$ be a lognormal random sample LN(1,1), of size $n_0=100$. 
The largest observed
data point was $\max(\bfX_0)=32.36495$, while $T=59.75377$. Hence, 
$\max(\bfX_0)=32.36495 < < T$. We wish to estimate the probability of exceeding $T$.
That is, we wish to estimate the tail probability $p=P(X > T)$, which in the present case is
$p=0.001$.

As in (\ref{RepeatedFusion}),
we fused $\bfX_0$ repeatedly
with 10,000 computer generated Unif(0,80) samples of size $n_1=n_0=100$. The iterative method
produces ``D=Down" and ``U=Up" sequences. When the ``Down" transitions to ``Up" we know we are very close to the true $p$. In the present example we have:\\
\\
D 0.001199466, D 0.001099466, D 0.000999465, D 0.000999465,\\ 
D 0.000999465, D 0.000999465, U 0.000999465, U 0.000999465.\\
\\
Here the transition from Down to Up occurs at $0.000999465$, which can be taken as point estimate for
$p=0.001$. In this example the error is quite small. 
%of order $\mathcal{O}(10^{-07})$. 

This ``Down-Up" phenomenon, giving surprisingly precise estimates, 
has been observed numerous times across many different tail types as we shall demonstrate in 
Tables \ref{LN11 p=0.001} to \ref{Mercury_II  p=0.001} in 
Section \ref{Illustrations of an Iterative Process}, where 
we also return to the present example providing more details.

\section{ROSF and the B-Curve}
\label{ROSF and the B-Curve}

We are in pursuit of a small tail probability $p$.
It is shown how to construct a curve which contains with a high probability
a point whose ordinate is $p$.

Suppose $\bfX_0=(X_1,...,X_{n_0})$ is a reference sample 
from some reference distribution $g$, and that
we wish to estimate a small tail probability $p$ of that distribution. 
The $X_i \sim g$ could represent quantities such as earthquake magnitude,
radioactive
contamination, claim amount, financial return, poverty level, wealth, temperature, 
and so on, and the interest is in the tail probability
$p=P(X > T)$ for some relatively high threshold $T$.

Combining 
the reference sample $\bfX_0$ with a computer-generated sample $\bfX_1$ gives the
fused sample $(\bfX_0,\bfX_1)$. 
Then $\bfX_0$ can be fused
again with another independent computer generated sample $\bfX_2$ and we get another 
fused sample $(\bfX_0,\bfX_2)$,
and so on.
%($\bfX_1$ is used generically), 
All these computer-generated samples $\bfX_1$, $\bfX_2$,... are independent and are
generated in an identical manner, and all have the same size $n_1$.
We refer to these computer-generated samples as {\em fusion samples}. Observe that the fused 
or combined samples $(\bfX_0,\bfX_1)$, $(\bfX_0,\bfX_2)$,... all have size $n_0+n_1$.

Here is how B-curves are constructed. 
We fuse the given reference sample $\bfX_0$ with a computer-generated fusion sample 
$\bfX_1$ from $g_1$ and get 
in a certain way, described in the next section, 
a confidence interval  for the small tail probability $p$. Let 
$B_1$ denote the upper bound of that interval.
%Since $p$ is small we take the lower bound to be $0$, and compute the upper bound $B_1$.
We fuse the 
given reference sample $\bfX_0$ again with another artificial fusion sample $\bfX_2$ 
 from $g_1$
and get in the same manner 
another upper bound $B_2$ for $p$.  
This process is repeated many times to produce a long sequence of confidence intervals 
 and hence a long sequence of upper bounds $B_i$, $i=1,2,...$. 
Conditional on
$\bfX_0$, the sequence of upper bounds $B_1,B_2,...$ is then 
an independent and identically distributed  
sequence of random variables from some distribution $F_{B}$.
It is assumed that 
\begin{eqnarray}
P(B_1 > p) = 1 - F_{B}(p) > 0.
\label{P(B_1 > p)>0}
\end{eqnarray}
%%%%Lu's Theorem give $95\%$ coverage, in specified cases!!! 
%%%%But we are not sure about misspecified cases. Hence we need assumption 
%%%%(\ref{P(B_1 > p)>0})
Let $B_{(1)},B_{(2)},...,B_{(N)}$ be a sequence of order statistics from smallest to largest.
Then, as $N \rightarrow \infty$, $B_{(1)}$ decreases and $B_{(N)}$ increases. Hence, as
the number of fusions increases
the plot consisting of the pairs 
\begin{eqnarray}
(1,B_{(1)}),(2,B_{(2)}),...,(N,B_{(N)})
\label{B-Curve}
\end{eqnarray}
contains a point whose ordinate is $p$ with 
probability approaching 1.
It follows that as $N \rightarrow \infty$, there is a $B_{(j)}$ which essentially coincides with $p$.
The plot of points consisting of the pairs $(j,B_{(j)})$ in (\ref{B-Curve})
%of ordered upper bounds $B_{(1)},B_{(2)},...,B_{(n)}$ 
is referred to as the {\em B-curve}.

Typical B-curves corresponding to the tail probability
$p=P(X > T)=0.001$ for various reference samples $\bfX_0$ from the
indicated distributions or data are shown in Figure 1. 
Notice that to get $p=0.001$,
in each case the threshold $T$ must change accordingly, and that in each plot there is a $B_{(j)}$
nearest or closest to $p=0.001$. The curves were obtained from 10,000 fusions of $\bfX_0$ 
with uniform samples with support exceeding $T$. Clearly, in all cases $B_{(1)} < p < B_{(10,000)}$.

Figure \ref{B-Curves f_2_7} shows B-curves from the f(2,7) distribution for various 
$\max(\bfX_0)$ where $T=21.689$. The point ``$\bullet$" moves to the left as $\max(\bfX_0)$ increases
relative to $T=21.689$. We can see that in each plot there is a $B_{(j)}$
nearest or closest to $p=0.001$.
As before, the curves were obtained from 10,000 fusions of $\bfX_0$ 
with uniform samples with support exceeding $T$, and in all cases
$B_{(1)} < p < B_{(10,000)}$.

\begin{figure}[htbp]
\begin{center}
\includegraphics[height=6.2cm,angle=-90]{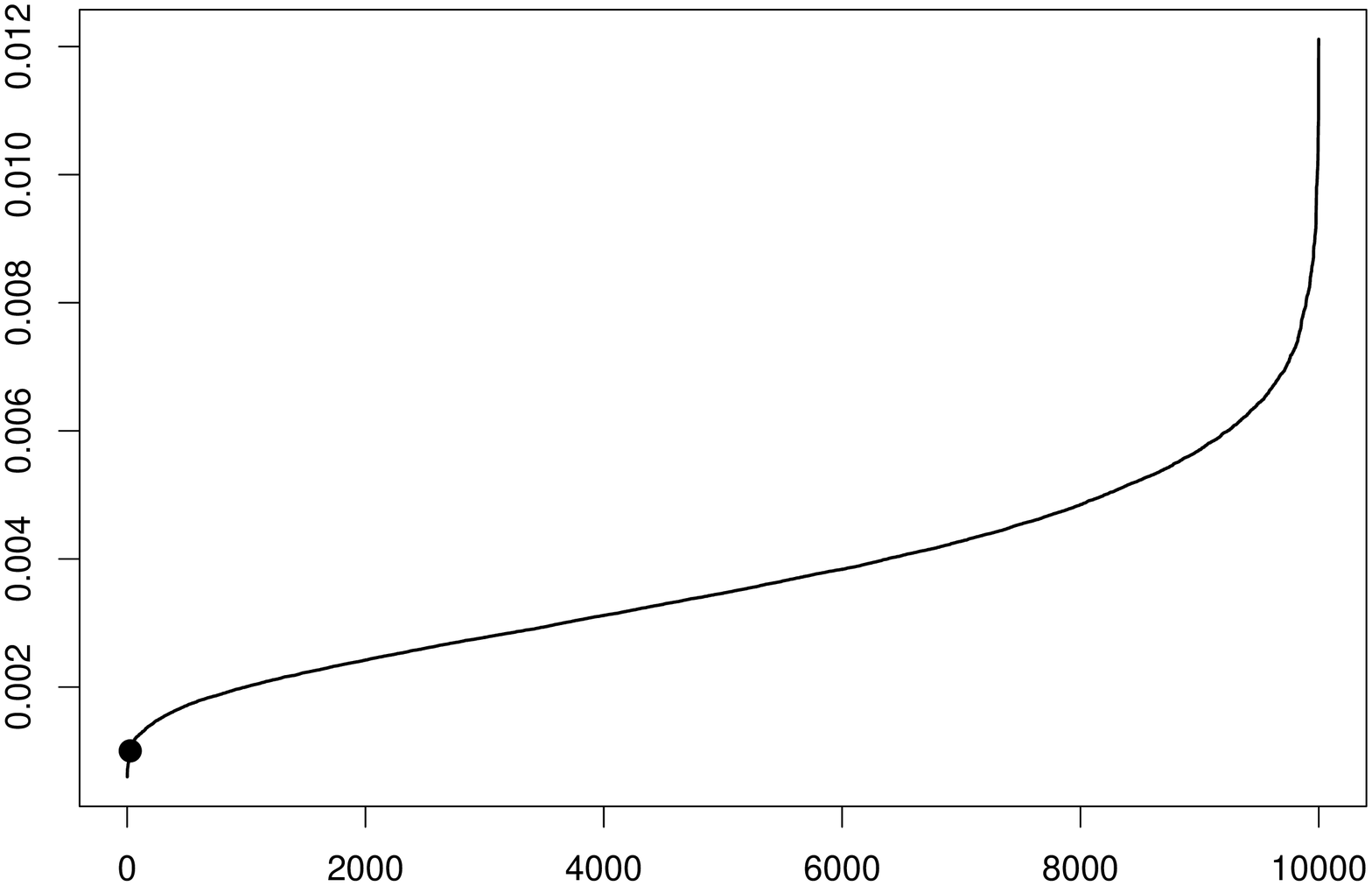}
\includegraphics[height=6.2cm,angle=-90]{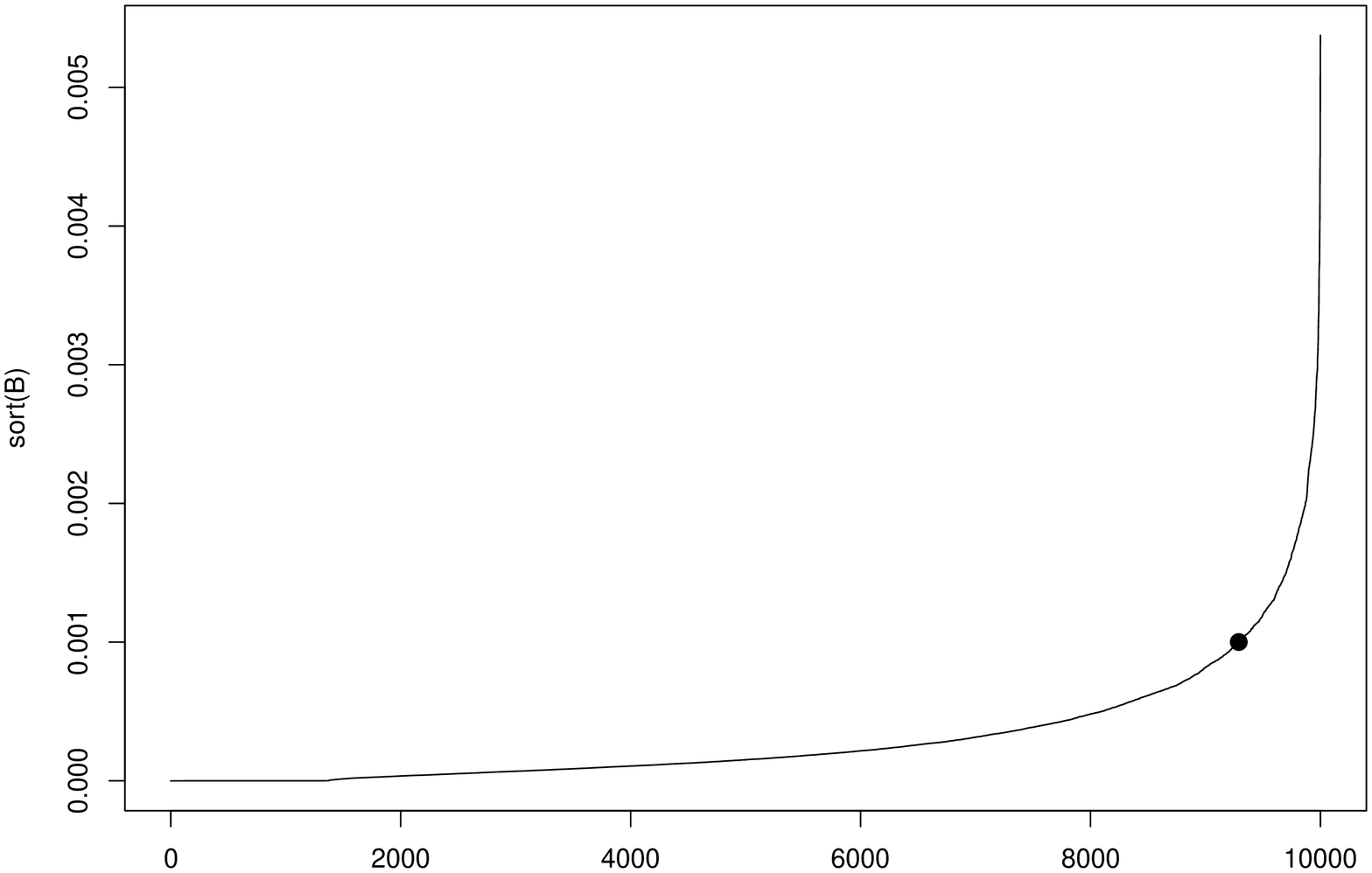}\\
\includegraphics[height=6.2cm,angle=-90]{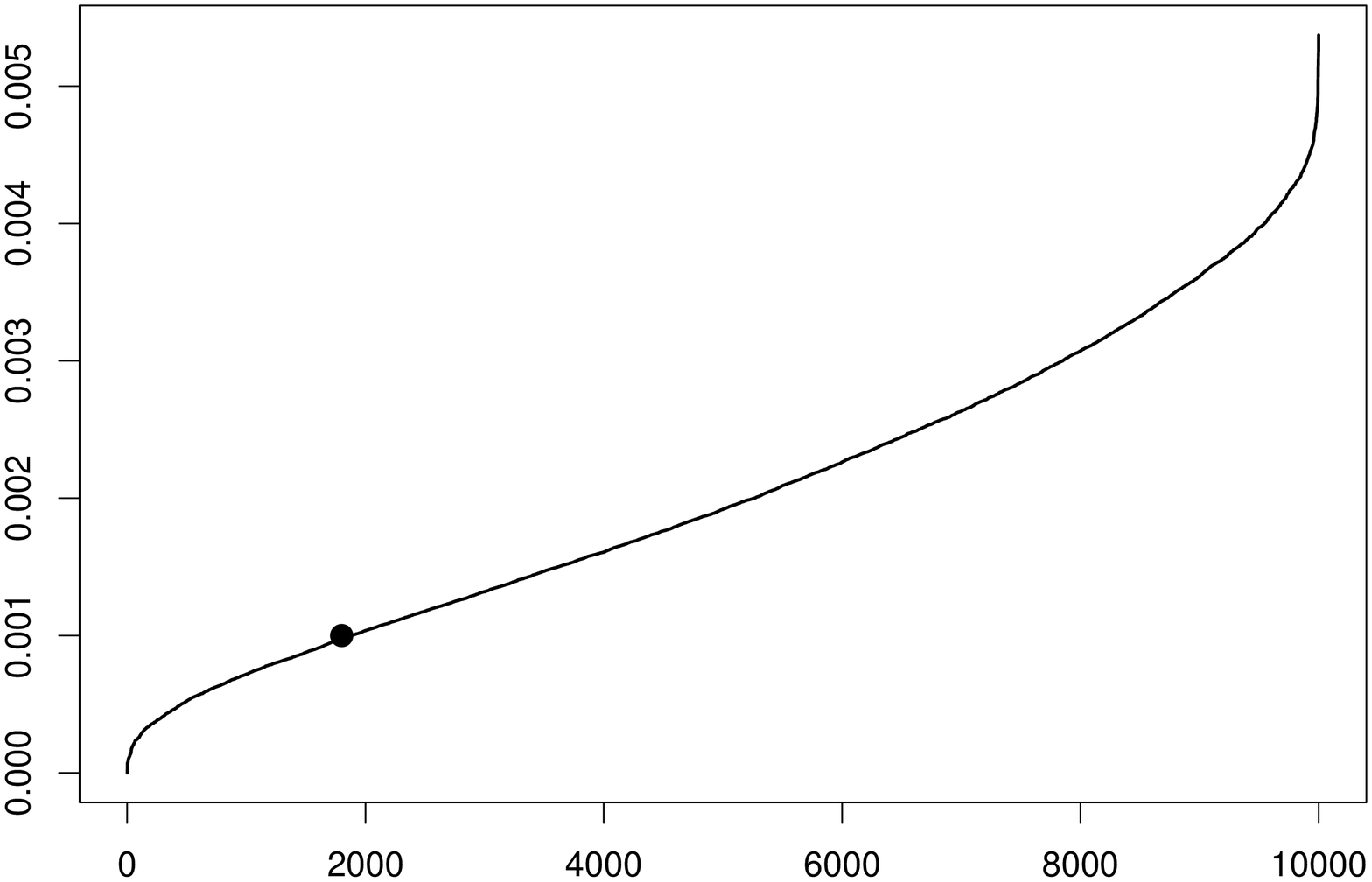}
\includegraphics[height=6.2cm,angle=-90]{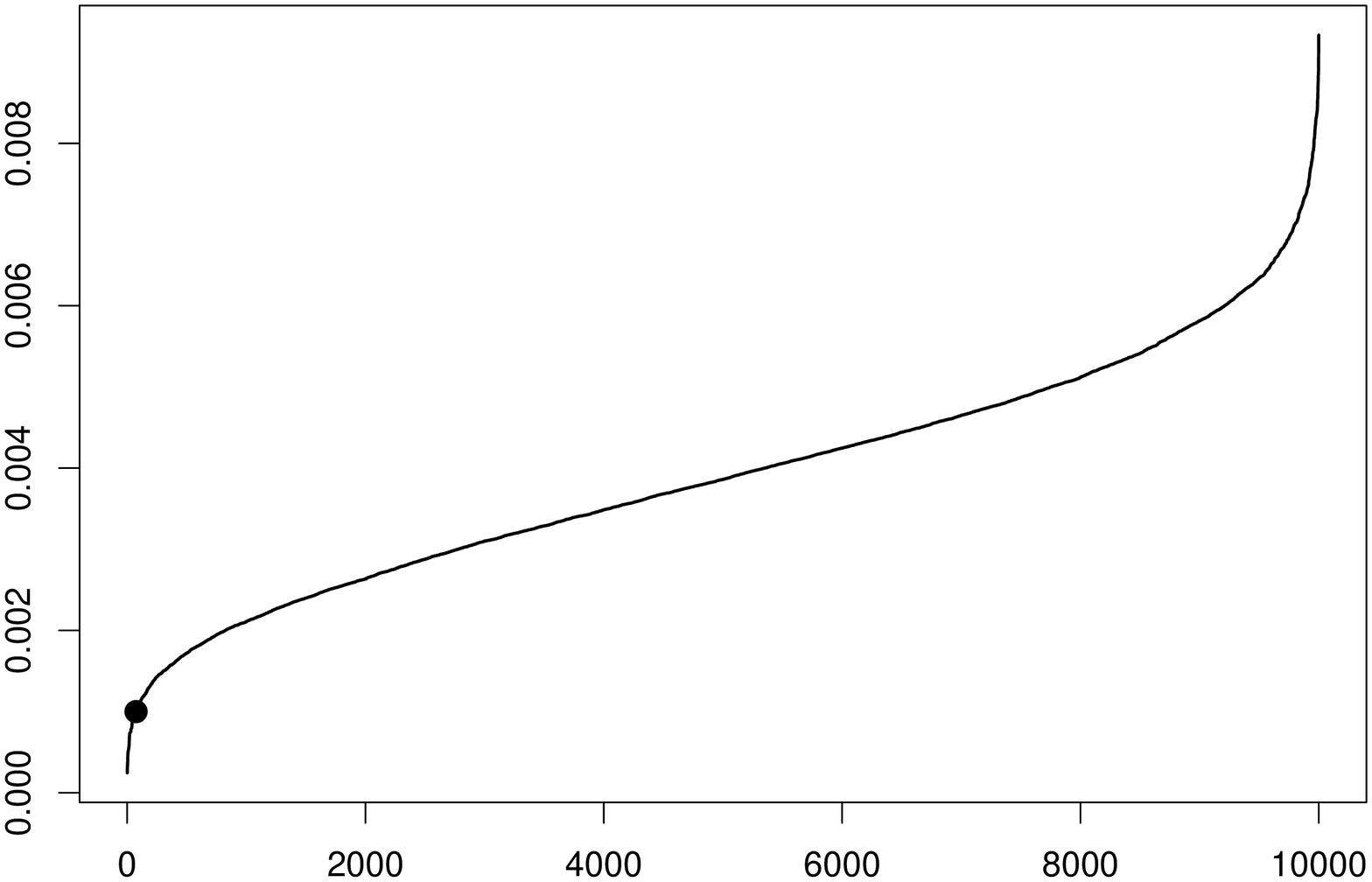}
\caption{Typical B-Curves from $B_{(1)},...,B_{(10,000)}$
containing a point corresponding to $p=0.001$. 
%toward the bottom of the curves. 
Clockwise from top left: 
Gamma(1,0.01), LN(1,1), Lead exposure, Mercury. 
$T$=690.7755, 59.7538, 25.00, 22.41, respectively, 
$n_0=n_1=100$. In all cases the fusion samples are uniform
with support exceeding $T$.}
%Histograms representing the distributions are shown in Figure 2.}
\end{center}
\label{B-Curves}
\end{figure}

\begin{figure}[htbp]
\begin{center}
\includegraphics[height=6.2cm,angle=-90]
{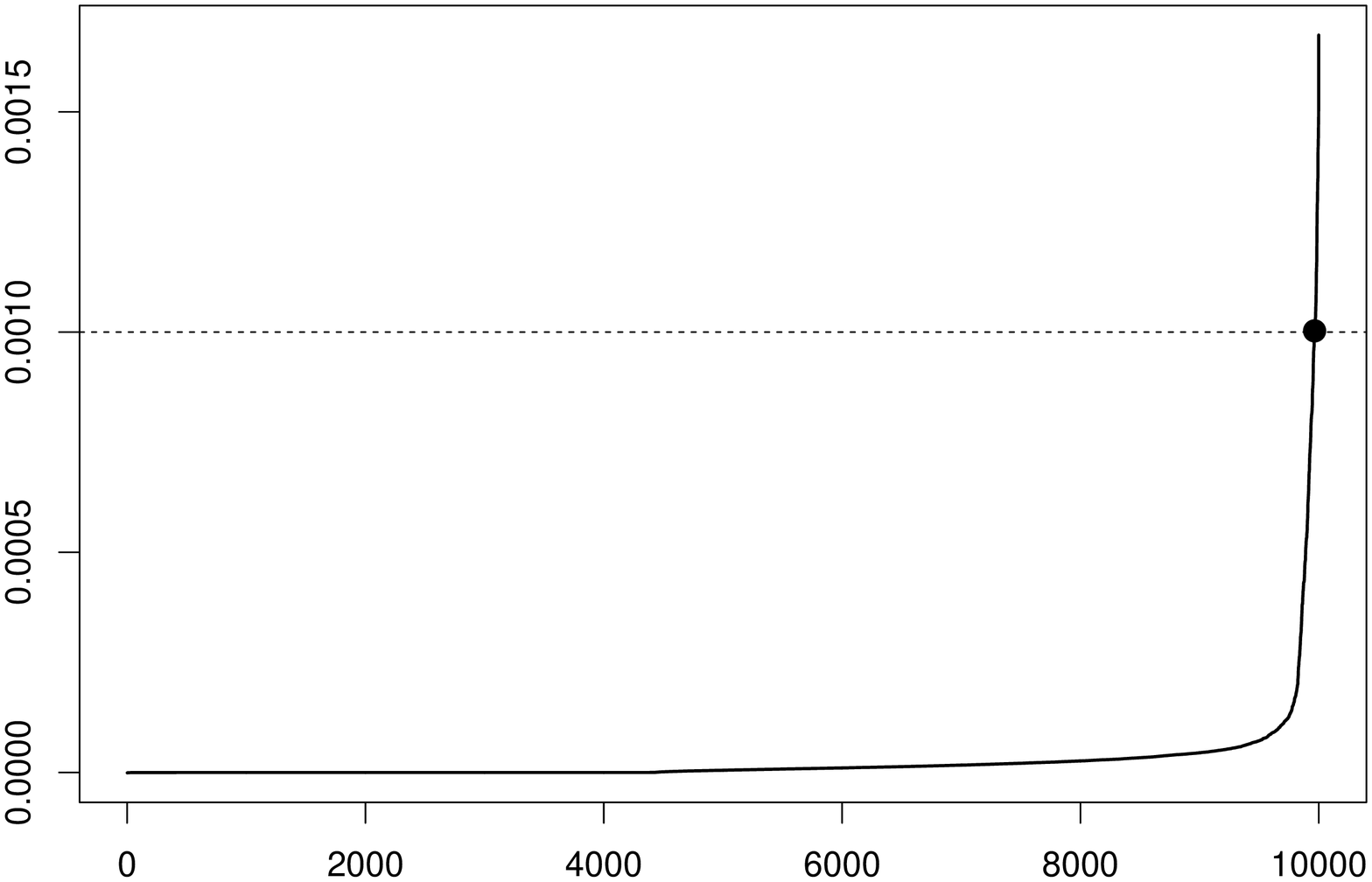}
\includegraphics[height=6.2cm,angle=-90]
{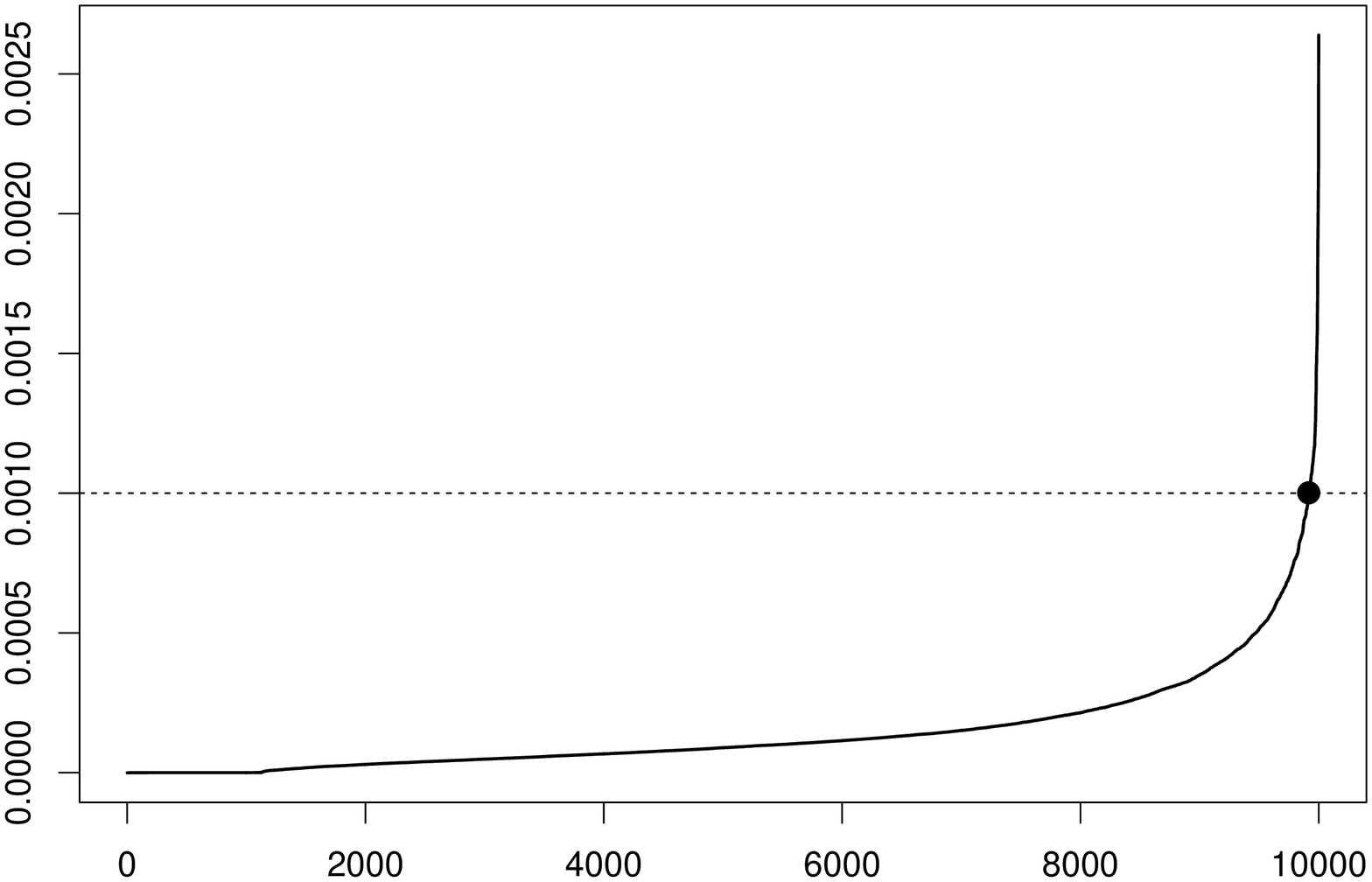}\\
\includegraphics[height=6.2cm,angle=-90]
{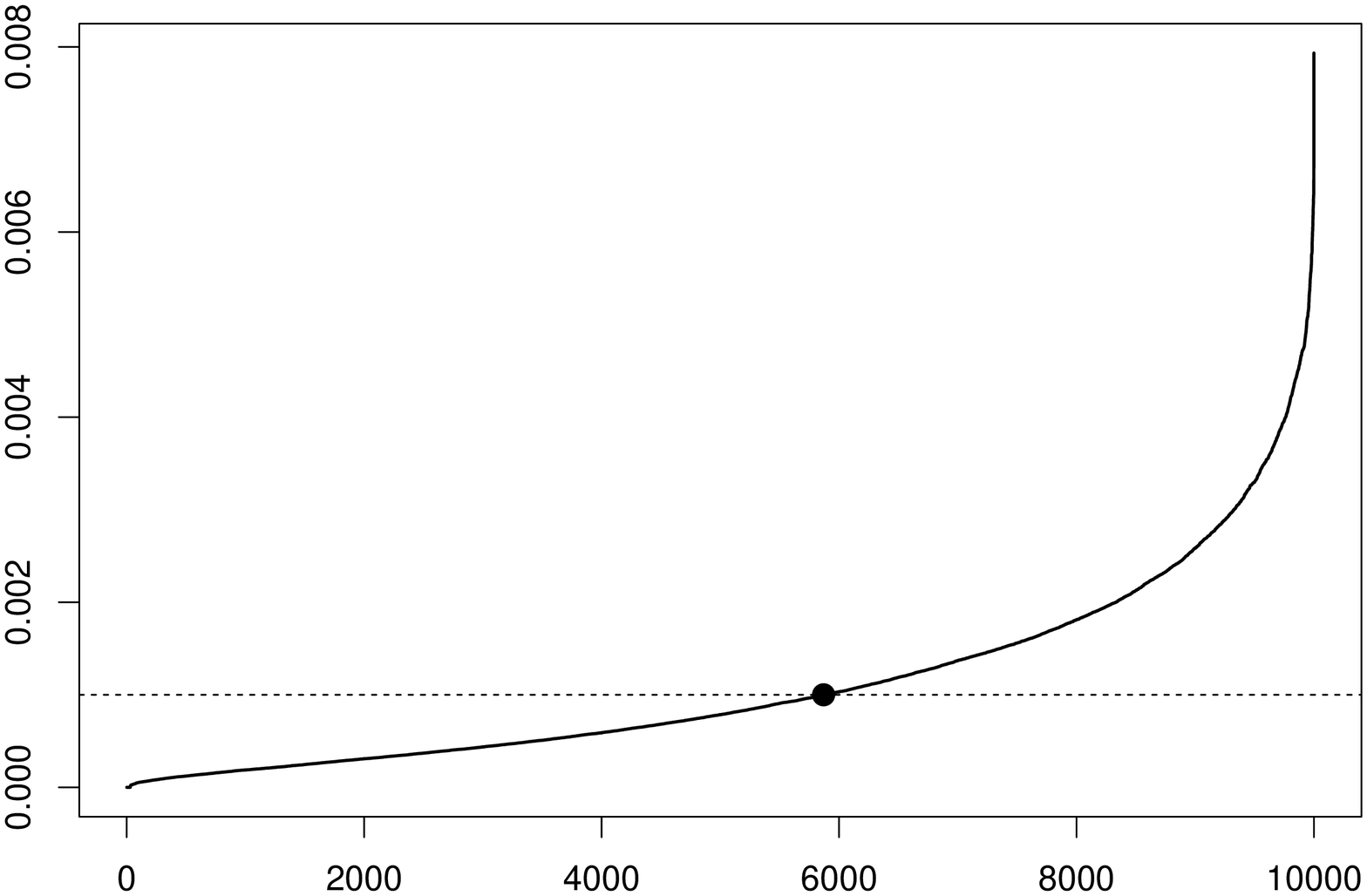}
\includegraphics[height=6.2cm,angle=-90]
{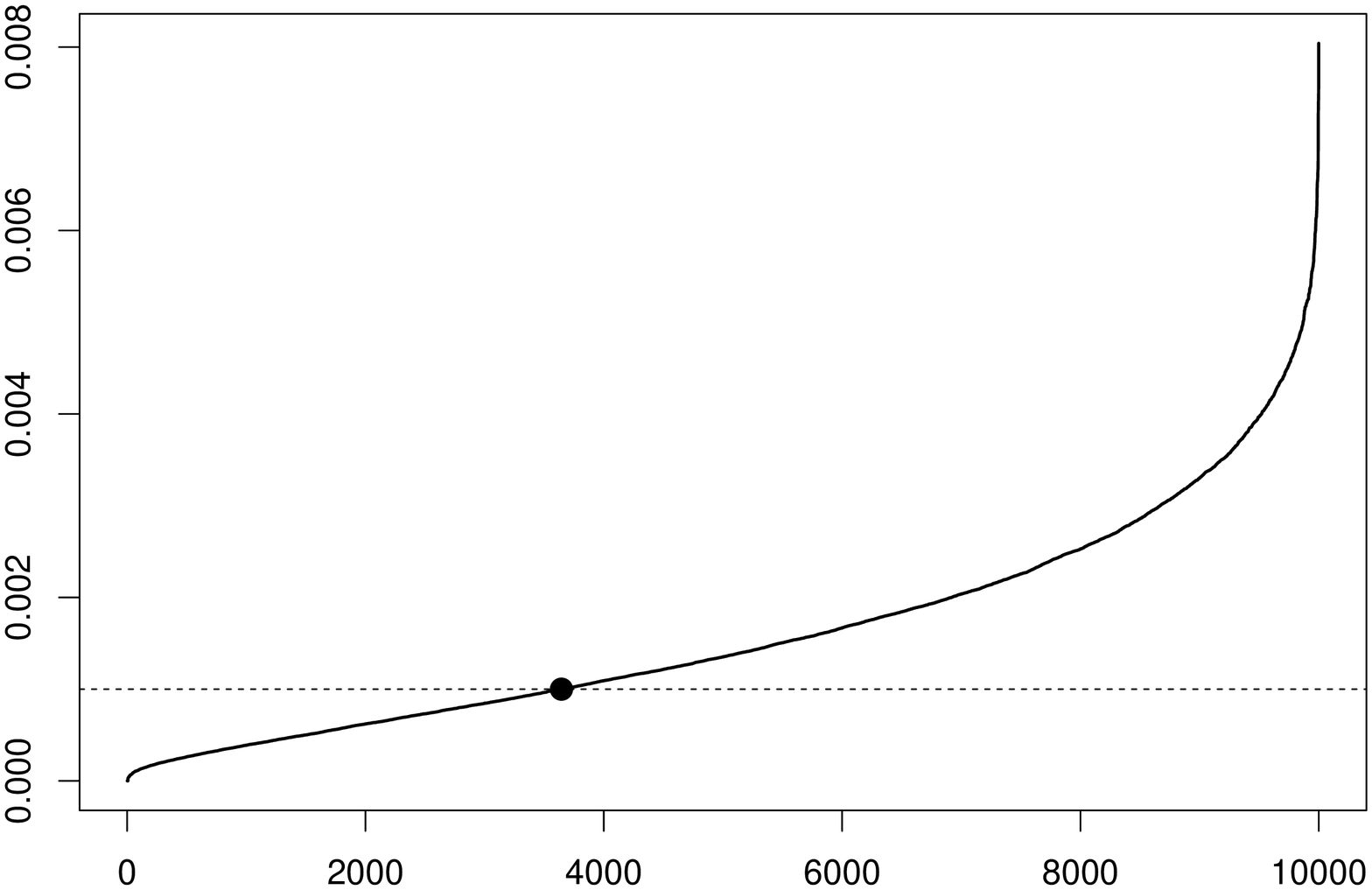}\\
\includegraphics[height=6.2cm,angle=-90]
{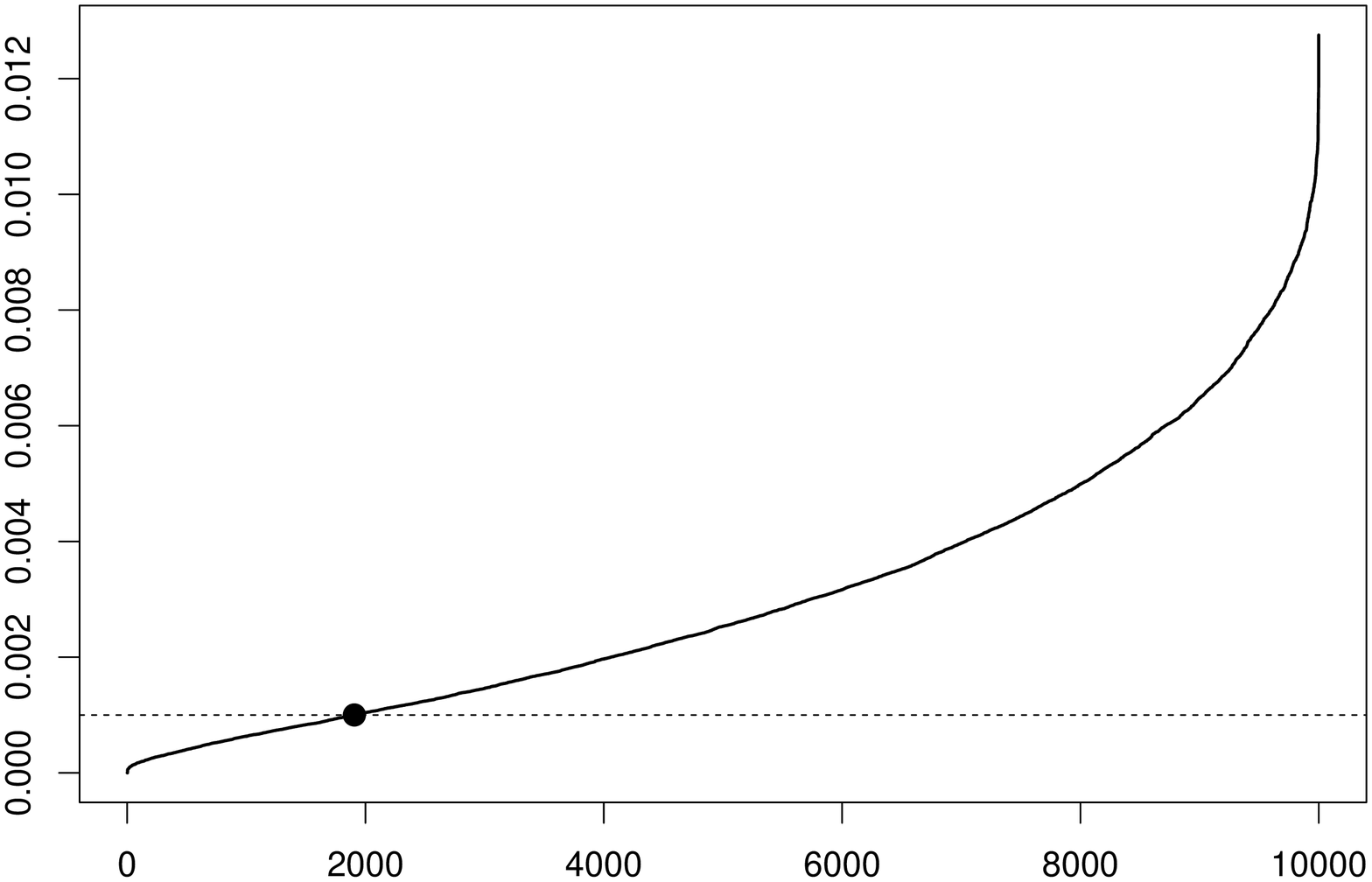}
\includegraphics[height=6.2cm,angle=-90]
{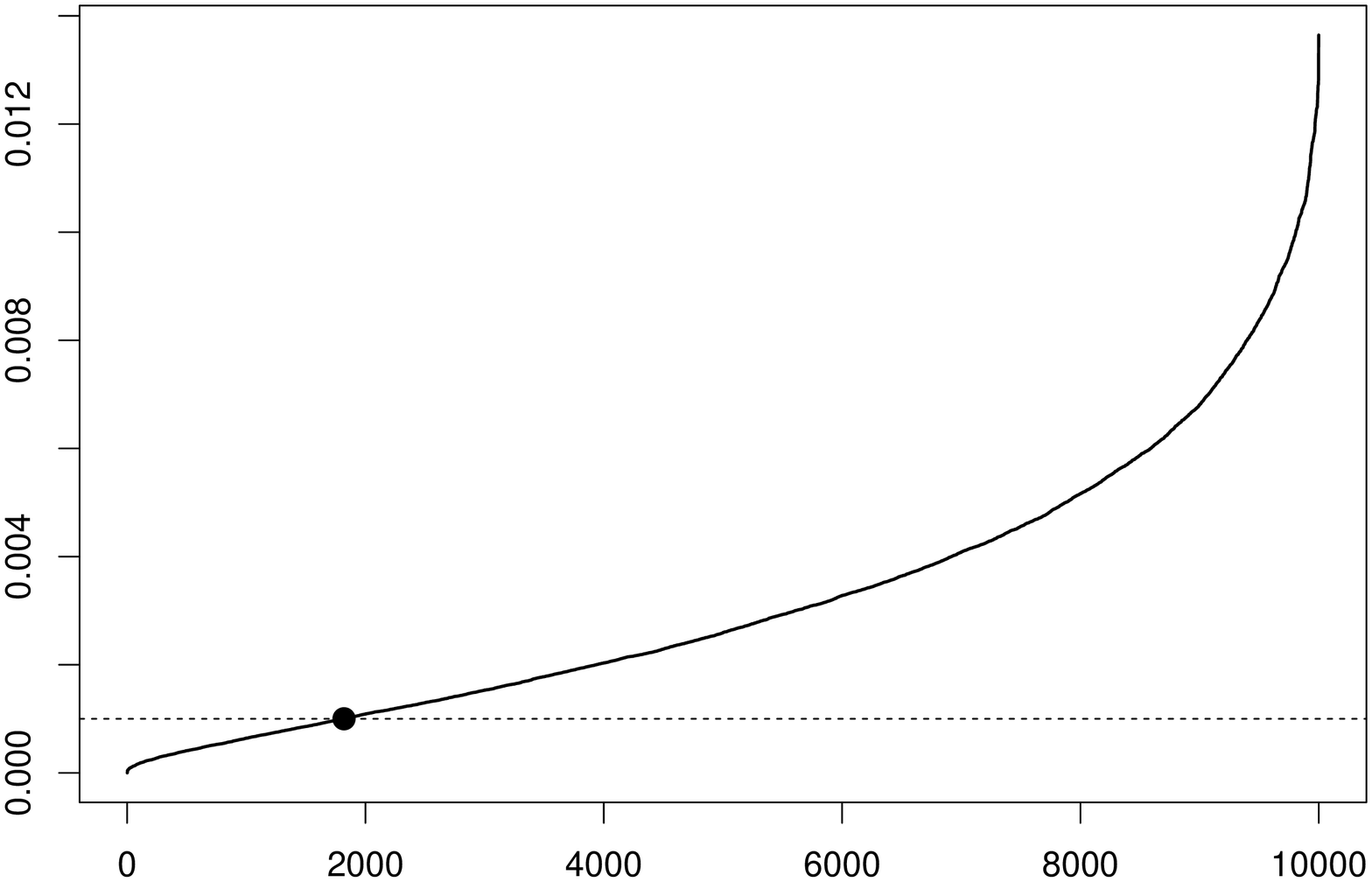}
\caption{B-Curves, 10,000 B's, from  f(2,7).
$n_0=n_1=100$, $p=0.001$,
$T=21.689$. 
$\max(\bfX_0)$ values: Top left 4.111466, top right 6.86562. 
Middle left 8.631132, middle right 10.18021. 
Bottom left 17.26258, bottom right 19.95937.
The point ``$\bullet$" moves to the left as $\max(\bfX_0)$ increases
relative to $T=21.689$. The fusion samples are uniform
with support covering $T$.}
%}
\end{center}
\label{B-Curves f_2_7}
\end{figure}

%\begin{figure}[htbp]
%\begin{center}
%\includegraphics[height=6.2cm,angle=-90]{Hist_Gamma_1_0.01.ps}
%\includegraphics[height=6.2cm,angle=-90]{Hist_LN_1_1.ps}\\
%\includegraphics[height=6.2cm,angle=-90]{Hist_Lead_Intake.ps}
%\includegraphics[height=6.2cm,angle=-90]{Hist_Mercury.ps}
%\caption{Histograms representing distributions with long right tails.
%The lead intake data are discussed in Kedem et al. (2016).
%The mercury data source is 
%NOAA's National Status and Trends Data
%\tt{https://products.coastalscience.noaa.gov/nsandt\_data/data.aspx}
%}
%\end{center}
%\label{Dist long tails}
%\end{figure}

%https://products.coastalscience.noaa.gov/nsandt_data/data.aspx

%The question then is how to find that $B_{(j)}$ to ``capture" $p$ on the B-curve. 
%We outline next an equivalent way of doing that. 
%{\em It is based on the fact that we can
%repeat the fusions indefinitely and hence obtain the distribution of the $B$'s.} 

A key fact of the present approach is that since the fusions can be repeated indefinitely,
we  can approximate the distribution of the $B$
upper bounds arbitrarily closely.

Let $\hat{F}_{B}$ be the empirical distribution obtained from the sequence of upper bounds $B_1,B_2,...,B_N$.
Then from the Glivenko-Cantelli 
Theorem, $\hat{F}_{B}$ converges to $F_{B}$ almost surely uniformly as
$N$ increases. 
Since the number of fusions can be as large as we wish, {\em our key idea},
$F_B$ is known for all practical purposes.

%We note that (\ref{P(B_(N) > p)}) implies that as $N \rightarrow \infty$,
%\begin{eqnarray}
%P(B_{(N)} > p) \rightarrow 1.
%\label{P(max_Bi > p)}
%\end{eqnarray}
%Thus, as the number of $B_i$ grows,
%the confidence interval $[0,\max{B_i}]$ covers $p$ with a high probability close to $1$;
%see Figure \ref{B-Curves}.
%Statement (\ref{CI}) is a refinment of this fact.

\section{Getting Upper Bounds by Data Fusion}
\label{Getting Upper Bounds by Data Fusion}

%Clearly, the preceding argument is quite general, and 
%the effectiveness of the procedure will depend on the quality of the $[0, B_i]$ confidence %intervals. 
%these confidence intervals, which amounts to
This section describes a particular way of generating upper bounds for tail probabilities $p$
by   data fusion of the real $\bfX_0$
and additional computer-generated data (``augmented reality" as it were)
under the density ratio model defined in (\ref{density-ratio}) below. 
The upper bounds are needed in order to generate B-curves.

In general, by
``fusion" or ``data fusion" we mean  the combined data from $m+1$ sources
where each source is governed by a probability
distribution. In the spirit of augmented reality, 
computer algorithms which generate random data
are perfectly legitimate data sources.
Using the combined data, semiparametric statistical inference can be ensued 
under the density ratio model assumption (Kedem, et al. 2017, Lu 2007, Qin and Zhang 1997).

Assume that the reference random sample $\bfX_0$ of size $n_0$ follows an
unknown reference distribution with probability 
density $g$, and let $G$ be the corresponding cumulative distribution function (cdf).

Let 
$$\bfX_1,...,\bfX_m,$$ be additional
computer-generated 
random samples  where $\bfX_j \sim g_j,G_j$, with size $n_j$, $j=1,...,m$.
As in the Appendix, for now $m \ge 1$ but later we specialize to $m=1$ only as in 
(\ref{RepeatedFusion}).
The augmentation of $m+1$ samples
\begin{eqnarray}
\bft = (t_1,\ldots,t_n) = (\bfX_0,\bfX_1,\ldots, \bfX_m),
\end{eqnarray} 
of size $n_0 + n_1 + \cdots + n_m$ gives the fused data. 
%We shall assume the 
The density ratio model stipulates that
%(Qin and Zhang 1997, Lu 1997) 
\begin{eqnarray}
\frac{g_j(x)}{g(x)} = \exp(\alpha_j + \bfbeta_j' \bfh(x)), \quad j=1,\ldots,m,
\label{density-ratio}
\end{eqnarray}
where $\bfbeta_j$ is an 
$r \times 1$ 
parameter vector, $\alpha_j$ is
a scalar parameter, and $\bfh(x)$ is an 
$r \times 1$ 
vector valued
distortion or tilt function.
Clearly,
to generate the $\bfX_j$ samples we must know the corresponding $g_j$. However,
beyond the generating process, we do not make use of this knowledge.
Thus, by our estimation procedure,
none of the probability densities $g,g_1,...,g_m$ and the corresponding $G_j$'s, 
and none of the parameters $\alpha$'s and 
$\bfbeta$'s are assumed known, but, strictly speaking, the so called tilt function
 $\bfh$ must be a known function. 
However, in the present application  the requirement of 
a known $\bfh$ 
is apparently mitigated as accentuated by 
assumption (\ref{P(B_1 > p)>0}) above, which may hold for 
misspecified $\bfh$,
and by our examples with many different tail types.

Since all the probability distributions are connected by the density ratio model 
(\ref{density-ratio}), each distribution pair $g_j,G_j$ is estimated from the entire
fused data $\bft$ and not just from $\bfX_j$ only. The same holds for the reference
pair $g,G$.
Thus, for
example, the reference $G$ is estimated from the entire 
fused data $\bft$ with $n_0+n_1+\cdots +n_m$ observations and not only from 
the reference sample $\bfX_0$ with $n_0$ observations.

Under the density ratio model (\ref{density-ratio}), 
the maximum likelihood estimate of $G(x)$ based on the fused data $\bft$
is given in (\ref{G-hat}) in Section A.1 in the Appendix,
along with its asymptotic distribution described in Theorem
\ref{Theorem-weak-convergence-G-hat-G}. From the theorem we obtain 
confidence intervals  for $p=1-G(T)$ for any threshold $T$ 
using (\ref{SP Lu CI-pointwise for 1-G(T)}).
%(\ref{SP Lu CI-pointwise}). 
In addition, 
from (\ref{G-hat}) we get the point estimates for $p$
% $\hat{p}=1-\hat{G}(T)$
as well. It is different than the one obtained by the ``Down-Up" idea,
and is not used in the paper, as in many cases it underestimates $p$.

%Our strategy then is to obtain by ROSF 
%for some $\bfh$ 
%numerous upper bounds $B_i$ for $p=1-G(T)$ , $i=1,2,3,...$,
%for a
%relatively large $T$  using  (\ref{SP Lu CI-pointwise for 1-G(T)}) in the Appendix,
%assuming the density ratio model holds to a reasonable degree. 
%Conditional on $\bfX_0$,
%the upper bounds $B_i$ are  
%independent and identically distributed  
%random variables from some distribution $F_{B}$. 

%Under assumption (\ref{P(B_1 > p)>0}) $P(B_1 > p) > 0$.

%Obviously, the density ratio model per se need not hold, and even if it does for some 
%tilt function $\bfh$, 
%the validity or goodness of any choice of $\bfh$ is uncertain. Furthermore, 
%if  $\max(\bfX_0)$ is much smaller
%than the threshold $T$ then $\hat{p}$ from (\ref{G-hat}) could be much too small.

Our data analysis in Section  4 regarding many different tail types, and additional examples 
in Kedem et al. (2017),
indicate that 
%However, 
for the implementation of ROSF, the density ratio model need not hold precisely,
%any reasonable choice of $\bfh$ suffices 
%{\em as long as  assumption (\ref{P(B_1 > p)>0}) holds}, which is a mild requirement.
%Experience shows 
and that the ``gamma tilt" 
$\bfh(x)=(x,\log x)$ is a sensible choice for skewed data. 
%similar to those used 
%in Section \ref{Capturing a Point on the B-Curve}.
%Figure 2.
%Similarly, the ``lognormal tilt" $\bfh(x)=(\log x,(\log x)^2)$ is another possible choice.
%and take the lower bounds as $0$. 
%This is the ``certain way" referred to in the previous section by which 
%In this way we obtain numerous $[0,B_i]$ confidence intervals.
%When assumption (\ref{P(B_1 > p)>0}) holds, 
In that case,
many of the $B_i$ obtained from (\ref{SP Lu CI-pointwise for 1-G(T)})
will be
greater than $p$ as their number increases,
but some will not.
Hence, the 
{\em ordered} $B_{(i)}$ engulf or surround $p$ with probability approaching one 
as the number of fusions increases. 
That is,
as the number of fusions increases, the set of pairs $(j,B_{(j)})$ engulfs the 
desired point on the B-curve with probability approaching one. 
This is
illustrated in Figures 1 and 2 with 10,000 fusions.

%That is, 
%with a large number of fusions
%the ordered $B_{(j)}$
%engulf $p$ with a high probability.
This, in general, cannot be said about 
the ordered $\hat{p}$'s from (\ref{G-hat})
unless the number of fusions is exceedingly large. Hence, we shall not
use $\hat{p}=1-\hat{G}(T)$ from  (\ref{G-hat}). Instead we estimate $p$ from the iterative 
process described earlier and in Section \ref{Capturing a Point on the B-Curve}.

%Thus, $[0, B_1]$ is obtained
%from the first fusion of $\bfX_0$ with a  set of $m$ computer-generated samples.
%Then $[0,B_2]$ is obtained by fusing $\bfX_0$ again but with a {\em different} independent
%set of $m$ computer-generated samples, and so on. 
%From each fusion we obtain a point estimate $\hat{p}=1-\hat{G}(T)$
%using (\ref{G-hat}) and an upper bound $B_i$ from (\ref{SP Lu CI-pointwise for 1-G(T)}).
%Since this fusion process is repeated numerous times, we obtain both numerous 
%point estimates $\hat{p}$'s and numerous
%upper bounds $B_i$'s. 

%See 
%Section \ref{Lognormal(0,1)} for a case where the $\hat{p}$'s from (\ref{SP Lu CI-pointwise for 1-G(T)})
%are too small. 

%We shall refer to this interval estimation 
%method as SP, for semiparametric.

In this paper $m=1$ only, and the fusion samples are uniform random samples
supported over a wide range which covers $T$.
% that is,  $\max(\bfX_1) > T$. But why uniform? First,  
The reason for uniform samples is that
when the density ratio model holds for some $g$
and $g_1$, then it also holds approximately by taking $g_1$ as a uniform density supported over 
a sufficiently wide range.

%Second, and more to the point,
%ROSF requires only the mild assumption (\ref{P(B_1 > p)>0}). 
%Experience shows that assumption (\ref{P(B_1 > p)>0}) holds well when fusing  
%$\bfX_0$ with uniform samples
%using the tilt function $h(x) = (x, \log x)$
%{\em across a wide range of tail types}.
%Evidently,
%the B-curves used in this paper provide further support for the validity of assumption 
%(\ref{P(B_1 > p)>0}).

%To summarize, 
%under assumption (\ref{P(B_1 > p)>0}), 
%the B-curves are constructed from a large number of ordered upper bounds $B_{(j)}$ from
%(\ref{SP Lu CI-pointwise for 1-G(T)}) for $p=P(X > T)$. 
%Recall that the $B_{(j)}$ upper bounds are 
%obtained from 
%a large number of fusions of $\bfX_0$ with random uniform samples $\bfX_1$, $\bfX_2$,
%$\bfX_3$,...,
%where the upper limit of the uniform distribution exceeds $T$. Throughout the paper,  
%$\max(\bfX_0) < T$ and $h=(x,\log x)$.

To summarize, numerous examples with skewed data suggest 
that the confidence intervals (\ref{SP Lu CI-pointwise for 1-G(T)})  are still useful 
in conjunction with $\bfh(x)=(x,\log x)$ even when the density ratio model does not hold in a strict sense.
In that case,
%assuming that the density ratio model holds reasonably,
the reference sample
$\bfX_0$ is fused repeatedly with identically distributed
independent random uniform samples $\bfX_1$, $\bfX_2$,..., as in (\ref{RepeatedFusion}),
where the upper limit of the uniform support exceeds $T$.
Repeated fusion gives upper bounds $B_1,B_2,...$ 
for $p=P(X > T)$ using (\ref{SP Lu CI-pointwise for 1-G(T)}). 
%Assuming (\ref{P(B_1 > p)>0}), 
Conditional on $\bfX_0$,
the upper bounds $B_i$ are  
independent and identically distributed  
random variables from some distribution $F_{B}$. 
The B-curves are constructed from a large number of pairs $(j,B_{(j)})$, where the
$B_{(j)}$ are the ordered upper bounds.

%In Kedem et al. (2016) it has been demonstrated that in many cases, 
%when g is skewed, not necessarily gamma, 
%and particularly when the tail is sufficiently thin, 
%then fusing the given reference sample
%$\bfX_0$ repeatedly with computer-generated uniform random samples $\bfX_1$ (i.e.
%ROSF) with uniform density $g_1$, and using the tilt function $h(x) = (x, \log x)$,
%provides surprisingly precise confidence intervals for small exceedance probabilities,
%using only moderately large reference samples.
%Moreover, this has been the case across a fairly wide range of tail types.
%This will be demonstrated in the present paper too using B-curves.

\section{Capturing a Point on the B-Curve}
\label{Capturing a Point on the B-Curve}

Due to a large number of fusions $n$, 
 $F_B$ is known for all practical purposes and with probability close to 1
\begin{eqnarray}
B_{(1)} < p < B_{(n)}.
\label{B_1 LT p LT B_n}
\end{eqnarray}
%for a very large $n$. 
In general, even for $n=1,000$, $B_{(1,000)}$ is much larger than the true $p$
and $B_{(1)}$ is very close to $0$. The goal is to find $B_{(j)}$ close to $p$.

It follows, by the monotonicity of the B-curve
and (\ref{B_1 LT p LT B_n}), that as $j$ {\em decreases} (for example 
from $n=10,000$), the $B_{(j)}$ approach $p$ from above so that
there is a $B_{(j)}$ very close to $p$. Likewise, the $B_{(j)}$ can approach $p$ from below.
Thus, the B-curve establishes a relationship between
$j$ and $p$.

Another relationship between $j$ and $p$ is obtained 
from a basic fact about order statistics where it is known that
\begin{eqnarray}
P(B_{(j)} > p) = \sum_{k=0}^{j-1} {n \choose k} 
[F_B(p)]^k [1- F_B(p)]^{n-k}.
\label{P_B_j_GT_p}
\end{eqnarray}
%where 
%\begin{eqnarray}
%\sum_{k=0}^{j-1} {n \choose k} 
%[F_B(B_{(j)})]^k [1- F_B(B_{(j)})]^{n-k}
%\label{P_B_j_GT_B_j}
%\end{eqnarray}
%holds with a high probability. This provides an approximation for $j$. 
%With this $j$, going back to  (\ref{P_B_j_GT_p}),
%$p$ can be approximated from 
%Clearly, we wish to find a $B_{(j)}$ very close to $p$ where (\ref{P_B_j_GT_p}) holds with a 
%high probability, say 0.95.

%Consider now the solution $p_{j_1}$ to the equality 
%$$P(B_{(j_1)} > p_{j_1}) =0.95$$

Suppose now that the probabilities 
$$P(B_{(j_1)} > p_{j_1}), \ P(B_{(j_2)} > p_{j_2}),... .$$
are sufficiently high probabilities, and that from the B-curve we get the close approximations 
$$p_{j_1} \stackrel{.}{=} B_{(j_2)}, \ p_{j_2} \stackrel{.}{=} B_{(j_3)}... .$$
Then with a high probability we get a decreasing ``down" sequence  
$$B_{(j_1)} > B_{(j_2)} > B_{(j_3)} \cdots .$$ 
 Replacing the ``sufficiently high probabilities"  by
``sufficiently low probabilities",
then a dual argument leads to an increasing
``up" sequence
$$B_{(j'_1)} < B_{(j'_2)} < B_{(j'_3)} \cdots .$$
%as (\ref{P_B_j_GT_p}) is monotone decreasing,
Thus, when the probabilities are sufficiently high the $B_{(j_k)}$ decrease, and when the 
probabilities are
sufficiently low the $B_{(j_k)}$ increase. 
In particular, this ``Down-Up" phenomenon occurs in a neighborhood of the true $p$,
where a {\em transition or shift} occurs from ``down" to ``up" or vice versa,
resulting in a ``capture" of $p$. 
Thus, allowing for high and low probabilities by bounding (\ref{P_B_j_GT_p}) by a
sufficiently high probability, we have.\\
\\
%This leads to the following proposition regarding a neighborhood of the true $p$.\\
{\bf Proposition}:
{\em Assume that the samples size $n_0$ of $\bfX_0$ is large enough, and that
the number of fusions $n$ is sufficiently large so that $B_{(1)} < p < B_{(n)}$.
Consider 
the smallest $p_j \in (0,1)$ which satisfy the inequality
\begin{eqnarray}
P(B_{(j)} > p_j)=
\sum_{k=0}^{j-1} {n \choose k} 
[F_B(p_j)]^k [1- F_B(p_j)]^{n-k} \leq 0.95,
\label{P_B_j_GT_p_LT_0.95}
\end{eqnarray} 
where the $p_j$ are evaluated along appropriate numerical increments.
Then, 
%by the same dual argument, 
(\ref{P_B_j_GT_p_LT_0.95}) 
produces ``down" and ``up" sequences depending on the  $B_{(j)}$ relative to $p_j$.
In particular, in a neighborhood of the true tail probability $p$, 
%convergence occurs when the ``down" sequence changes course to an ``up" sequence. 
%In other words, 
with a high probability,
there are 
``down" sequences which converge from above and ``up" sequences which converge
from below 
to points close to $p$.}\\
%provides another relationship between $j$ and $p$. Note that
%if ``$>$" is used instead of ``$\leq$" in (\ref{P_B_j_GT_p_LT_0.95})
%then the solution of (\ref{P_B_j_GT_p_LT_0.95}) is $p=0$. This is so since
\\
This will be demonstrated copiously
across different tail types using an
approximation to (\ref{P_B_j_GT_p_LT_0.95}). 
We note that 
(\ref{P_B_j_GT_p}) is  a steep monotone decreasing step function 
so that
if ``$>$" is used instead of ``$\leq$" in (\ref{P_B_j_GT_p_LT_0.95})
then the solution of (\ref{P_B_j_GT_p_LT_0.95}) is $p=0$,
%of the type shown in Figures 3,4,
and that replacing $0.95$ by $0.99$ in (\ref{P_B_j_GT_p_LT_0.95}) gives similar results.

Iterating between these two monotone relationships, the B-curve and (\ref{P_B_j_GT_p_LT_0.95}),
is what was referred to earlier as 
the iterative method (IM). The iterative method provides our $p$ estimates.
%\ref{StepFunLN11_Max_x0_25.17781_}.
The iterations could start with a sufficiently large $j$ suggested by the B-curve, or,
alternatively with a sufficiently small $j$,  
until the down and up sequences 
converge to the same or very close points. The average of these points, or an approximation thereof, 
is our point estimate from the iterative process and it is
different than $\hat{p}$ obtained from (\ref{G-hat}) in the Appendix.

In general, 
 starting with any $j$,
convergence occurs by monotonicity and 
we keep getting the same point.

%As $n$ increases, a reasonable initial $j$ is that $j$ corresponding to the average of the $B$'s
%giving $\bar{B} \approx B_{(j)} > p$.

%With that $j\equiv j_1$ 
%we look for the smallest $p\equiv p_{j_{1}}$ satisfying (\ref{P_B_j_GT_p_LT_0.95}). Next 
%we find a $B_{(j_2)}$ on the B-curve closest to $p_{j_{1}}$. 
%This gives a new $j\equiv j_2$ and the previous steps
%are repeated 
%until convergence occurs and 
%we keep getting the same $p$. This is our point estimate from the iteration process and it is
%different than $\hat{p}$ obtained from (\ref{G-hat}) in the Appendix.

In symbols, with $B_{(j_k)}$'s from the B-curve,
and
$p_{(j_k)}$'s the smallest $p$'s
satisfying (\ref{P_B_j_GT_p_LT_0.95}) with $j=j_k$,
and $B_{(j_{k+1})}$ closest to $p_{(j_k)}$, $k=1,2,...$,
$$B_{(j_1)} \rightarrow p_{(j_1)} \rightarrow B_{(j_2)}  \rightarrow \cdots 
B_{(j_k)} \rightarrow p_{j_k} \rightarrow B_{(j_{k+1})} \rightarrow p_{j_k}
\rightarrow B_{(j_{k+1})} \rightarrow p_{j_k} \cdots$$
so that $p_{j_k}$ keeps giving the same $B_{(j_{k+1})}$ 
(and hence the same $j_{k+1}$)
and vice versa. This can be expressed more succinctly as,
$$j_1 \rightarrow p_{(j_1)} \rightarrow j_2  \rightarrow p_{(j_2)}
\rightarrow \cdots 
j_k \rightarrow p_{j_k} \rightarrow j_{k+1} \rightarrow p_{j_k}
\rightarrow j_{k+1} \rightarrow p_{j_k} \cdots$$

As will be illustrated 
in Section \ref{Illustrations of an Iterative Process},
under some computational conditions this iterative process results in a
 contraction in a neighborhood of the true $p$.
%Convergence occurs since the $B_{(j_{k})}$ and the corresponding $p_{j_{k}}$ are of
%the same order so that in a neighborhood of the true $p$ we basically have a contraction
%$$|B_{(j_k)} - B_{(j_k^{'})}| < q |p_{(j_k)}-p_{(j_k^{'})}|$$
%for $q < 1$.

%In a small neighborhood of the true $p$ the $B_{(j_k)}$ can move either up or down
%as we shall see time and again from the examples in Section  
%\ref{Illustrations of an Iterative Process}.

%For example in the lead example in Section \ref{Lead Example}.

%https://www.math.umd.edu/~petersd/466/fixedpoint.pdf

\subsection{Computational Considerations}

Computationally, the iterative process depends on $n$ and the increments of $p$ at which 
(\ref{P_B_j_GT_p_LT_0.95}) is evaluated.
In practice, 
due to computational limitations of large binomial coefficients the iteration is done
as follows. After $F_B$ is obtained from a large number of fusions, say $n=10,000$ fusions
(which give 10,000 $B$'s),
%this process of using the two equations (\ref{P_B_j_GT_B_j}) and (\ref{P_B_j_GT_p_LT_0.95})
%can be amended in several ways. For example, 
%we estimate $p$ 
%by iterating between an approximate
%B-curve and (\ref{P_B_j_GT_p_LT_0.95}) as follows. 
%we first sample  
then $1000$ $B$'s are sampled at random
from the obtained  $n=10,000$ $B$'s to get an approximate B-curve.
Next, the
binomial coefficients  ${n \choose k}$ are replaced by ${1000 \choose k}$.
We then iterate between
an approximate
B-curve and approximate (\ref{P_B_j_GT_p_LT_0.95}) with $n=1000$ 
as in 
\begin{eqnarray}
\sum_{k=0}^{j-1} {1000 \choose k} 
[F_B(p_j)]^k [1- F_B(p_j)]^{n-k} \leq 0.95
\label{P_B_j_GT_p_1000}
\end{eqnarray}
%(as in  \ref{P_B_j_GT_p_1000} below)
until a ``Down-Up" convergence occurs, 
in which case an estimate for
$p$ is obtained as the Down-Up shift point. 
The iterative process is illustrated in the next section.
This procedure can be repeated many times
by sampling repeatedly many different sets of $1000$ $B$'s  
to obtain many point estimates $\hat{p}$ from which interval estimates can then be
constructed, as  well as variance estimates. 
%For example, if we have 10 point estimates. We can define min to max of the point estimates 
%be the interval estimates. By obtaining more point estimates, 
%we can compute the coverage of the interval estimates. 

Running 10,000 fusions
takes about 5 minutes in R which translates into about 8 hours for 1,000,000 fusions.
In what follows the $p$-increments at which (\ref{P_B_j_GT_p_1000}) is evaluated 
are 0.0001 when $p=0.001$ and 0.000015 when $p=0.0001$.
%chosen roughly as $\mathcal{O}(\bar{B})$.
In all cases the maxima (minima) of the approximate
B-curves occurred at a point  larger (smaller) than the true $p$, as in (\ref{B_1 LT p LT B_n}).

\subsection{Illustrations of an Iterative Process}
\label{Illustrations of an Iterative Process}

%%%%%%%%% Provide min max B!!!

%In the following illustrations ROSF is used with a  gamma tilt function 
%while the data are not gamma distributed. That is, the tilt function is misspecified.
%As long as assumption (\ref{P(B_1 > p)>0}) holds, the requirement of  a 
%correctly specified tilt function
%is abated considerably. 
%The results in Tables \ref{LN11 p=0.001} -- \ref{Mercury_II  p=0.001}
%support this.

The Down-Up convergence results together with the number of iterations are summarized in 
 Tables \ref{Gamma_1_0.05 p=0.001} to \ref{Weibull.8_2 p=0.0001}
for $p=0.001$ 
and also for $p=0.0001$. Due to insufficiently large data sets, the Down-Up convergence results
for real data 
in Tables \ref{URX3TB_trichlorophenol_I  p=0.001} -- \ref{Mercury_II  p=0.001} do not deal with 
the smaller $p=0.0001$.
The cdf $F_B$ was obtained from 10,000 $B$'s
(the result of 10,000 fusions), and
each entry in the tables was obtained from a 
{\em different} sample of 1,000 $B$'s sampled at random from 10,000 $B$'s.

%That is, for each entry in the tables 
%we first sample 1000 $B$'s from 10,000 $B$'s to get an approximate B-curve,
%by ordering the 1,000 $B$-values, and then
%iterate between it and the smallest $p$, along the indicated $p$-increments 
%($0.0001$ or $0.000015$)
%such that (\ref{P_B_j_GT_p_1000}) holds. The number of iterations decreases as the $B_{(j)}$
%enter a neighborhood of the true $p$, and the shift from down to up occurs 
%approximately around the true $p$.

\subsubsection{Some Typical Down-Up Sequences}

It is instructive first to realize some typical Down-Up sequences.
In the first example $\bfX_0$ is a LN(1,1) sample where $\max(\bfX_0)=32.36495$. 
With $T=59.75377$ the true tail probability
to be estimated is $p=0.001$, using $n_0=n_1=100$ and $\bfh=(x,\log x)$.  
The generated fusion samples $\bfX_1$ are from Unif(0,80), $80 > T$, 
and $F_B$ was obtained from 10,000 fusions.\\
\\
Typical convergent Down-Up sequences $(j,p_j)$ are given next. Again, 
each sequence was derived from a different $B$-sample of size 1000 drawn from
10,000 $B$'s. More examples are given in Kedem et al. (2018).\\
\\
\noindent
{\bf Down:}
$900 \rightarrow 0.001799466 \rightarrow  867 \rightarrow  0.001599466 \rightarrow  837 
\rightarrow  0.001499466
\rightarrow   822 \rightarrow  0.001399466 \rightarrow  801 \rightarrow  0.001299466 \rightarrow  
 778 \rightarrow 0.001199466 \rightarrow  751 \rightarrow  0.001099466 \rightarrow  723 
\rightarrow 0.001099466 \cdots,$\\
\\
{\bf Down:}
$800 \rightarrow 0.001299466 \rightarrow 775 \rightarrow 0.001199466 \rightarrow 743 
\rightarrow 0.001099466 \rightarrow 712 \rightarrow  0.0009994658 \rightarrow 680 \rightarrow 
0.0009994658 \cdots.$\\
\\
{\bf Up:}
$680 \rightarrow 0.0009994658 \rightarrow 694 \rightarrow 0.0009994658 \cdots,$\\
\\
{\bf Up:}
$670 \rightarrow 0.0008994658 \rightarrow 675 \rightarrow 0.0009994658 \rightarrow 711 
\rightarrow 0.0009994658 \cdots$.\\
\\
In the next example 
$\bfX_0$ is a mercury sample (see Section \ref{Mercury}), drawn from a large population, 
where $\max(\bfX_0)=7.99$. 
With $T=22.41$ the true tail probability
to be estimated is $p=0.001088797$, using $n_0=n_1=100$ and $\bfh=(x,\log x)$.  
The generated fusion samples $\bfX_1$ are from Unif(0,50), $50 > T$, 
and $F_B$ was obtained from 10,000 fusions.\\
\\
Typical convergent Down-Up sequences $(j,p_j)$ are:\\
\\
{\bf Down:}
$600 \rightarrow 0.001299352 \rightarrow  563 \rightarrow 0.001199352 \rightarrow 526 
\rightarrow 0.001099352 \rightarrow 502 \rightarrow 0.0009993515 \rightarrow
475 \rightarrow 0.0009993515 \cdots,$\\
\\
{\bf Down:}
$550 \rightarrow 0.001199352 \rightarrow 542 \rightarrow 0.001099352 \rightarrow 509 
\rightarrow 0.0009993515 \rightarrow 479 \rightarrow 0.0009993515 \cdots,$\\
\\
{\bf Up:}
$490 \rightarrow 0.0009993515 \rightarrow 503 \rightarrow 0.0009993515 \cdots$,\\
\\
{\bf Up:}
$470 \rightarrow 0.0008993515 \rightarrow  476 \rightarrow 0.0009993515 \rightarrow 505 
\rightarrow 0.0009993515 \cdots.$\\

{\em
We note that the number of Down-Up iterations decreases in a neighborhood of the true $p$.
As seen 
from Tables \ref{Gamma_1_0.05 p=0.001} to \ref{Mercury_II  p=0.001} below, in many cases 
1 or 2 iterations in a neighborhood of $p$ suffice. This reduction can serve as a telltale sign that convergence took place.
}

We further note that the Gamma cases in Tables \ref{Gamma_1_0.05 p=0.001} and 
\ref{Gamma_1_0.05 p=0.0001} are nearly specified whereas this cannot be said about the cases in 
Tables \ref{LN11 p=0.001} to \ref{Mercury_II  p=0.001}. However, the tables portray a 
very similar picture for both
real and simulated data, for (nearly) specified or misspecified cases,
giving precision on the order of 
$10^{-5}$ or better for $p=0.001$ and order of $10^{-6}$ for $p=0.0001$,
where $n_0=n_1=100$. 
The results in the tables were obtained from 10,000 out of sample fusions, and
in all cases it has been observed that
$p \in (B_{(1)},B_{(10,000)})$ as also illustrated in Figures 1 and 2.
{\em Notably, as seen from Tables \ref{Gamma_1_0.05 p=0.001} to \ref{Mercury_II  p=0.001}, 
the Down-Up shift point is close to $p$.}

\subsubsection{Gamma(1,0.05)}

%Provide $p \in (B_{(1)},B_{(10,000)})= (0.0001872,0.0063686)$.

\begin{table}[htbp]
\centering
\caption{\scriptsize
${\bf p=0.001}$, $\bfX_0 \sim Gamma(1,0.05)$, $\bfX_1 \sim Unif(0,170)$, $\max(\bfX_0)=73.0467$,
$T=138.1551$, $n_0=n_1=100$, $h=(x,\log x)$, $p$-increment $0.0001$.}
\scriptsize{
\begin{tabular}{rcrl}
Starting $j$ & Convergence to & Iterations &  \\
\hline
1000  &  0.002887173      &    13    &      Down\\
 400  &  0.001487173      &     1    &      Down\\
 300  &  0.001287173      &     1    &      Down\\
 230  &  0.001187173      &     1    &      Down\\
 215  &  0.001087173      &     1    &      Down\\
 210  &  0.001087173      &     1    &      Down\\
 200  &  0.001087173      &     1    &      Up\\
 180  &  0.001087173      &     1    &      Up\\
 150  &  0.000987172      &     1    &      Up\\
 140  &  0.000987172      &     1    &      Up\\ 
\end{tabular}
}
\label{Gamma_1_0.05 p=0.001}
\end{table}
A sensible estimate of $p=0.001$ is the average from the last 7 entries which gives 
$\hat{p}=0.001072887$
with absolute error of $7.2887 \times 10^{-5}$.

\begin{table}[htbp]
\centering
\caption{\scriptsize
${\bf p=0.0001}$, $\bfX_0 \sim Gamma(1,0.05)$, $\bfX_1 \sim Unif(0,210)$, $\max(\bfX_0)=77.61753$,
$T=184.2068$, $n_0=n_1=100$, $h=(x,\log x)$, $p$-increment $0.000015$.}
\scriptsize{
\begin{tabular}{rcrl}
Starting $j$ & Convergence to & Iterations &  \\
\hline
888  & 0.0003439967     & 2     & Down\\
577  & 0.0001339967     & 4     & Down\\
450  & 0.0001189967     & 3     & Down\\
350  & 0.0001189967     & 1     & Down\\
310  & 0.0001039967     & 1     & Down\\
300  & 0.0001039967     & 1     & Down\\
290  & 0.0001039967     & 1     & Up\\
280  & 0.0001039967     & 1     & Up\\
270  & 0.0001039967     & 1     & Up\\
260  & 0.0001039967     & 1     & Up\\
\end{tabular}
}
\label{Gamma_1_0.05 p=0.0001}
\end{table}
A sensible estimate of $p=0.0001$ is the value in the last 6 entries which gives 
$\hat{p}=0.0001039967$
with absolute error of $3.9967 \times 10^{-6}$.

\newpage
\subsubsection{Lognormal(1,1)}

\begin{table}[htbp]
\centering
\caption{\scriptsize
${\bf p=0.001}$, $\bfX_0 \sim LN(1,1)$, $\bfX_1 \sim Unif(0,80)$, $\max(\bfX_0)=32.36495$,
$T=59.75377$, $n_0=n_1=100$, $h=(x,\log x)$, $p$-increment $0.0001$.}
\scriptsize{
\begin{tabular}{rcrl}
Starting $j$ & Convergence to & Iterations &  \\
\hline
1000 & 0.001199466 & 21 & Down\\
 950 & 0.001099466 & 13 & Down\\
%900 & 0.001099466 &  8 & Down\\
900 & 0.000999465 & 10 & Down\\
%900 & 0.000999465 & 10 & Down\\
%800 & 0.001099466 &  4 & Down\\
 800 & 0.000999465 &  5 & Down\\
 750 & 0.000999465 &  3 & Down\\
%750 & 0.001099466 &  2 & Down\\ 
 700 & 0.000999465 &  2 & Down\\ 
%680 & 0.000999465 &  2 & Up\\ 
 680 & 0.000999465 &  2 & Up\\
 680 & 0.000999465 &  2 & Up\\
 670 & 0.000999465 &  2 & Up\\
\end{tabular}
}
\label{LN11 p=0.001}
\end{table}
A sensible estimate of $p=0.001$ is the average from the last 6 entries 
which gives $\hat{p}=0.000999465$
with absolute error of $5.35 \times 10^{-7}$.

\begin{table}[htbp]
\centering
\caption{\scriptsize
${\bf p=0.0001}$, $\bfX_0 \sim LN(1,1)$, $\bfX_1 \sim Unif(0,130)$, $\max(\bfX_0)=44.82807$,
$T=112.058$, $n_0=n_1=100$, $h=(x,\log x)$, $p$-increment $0.000015$.}
\scriptsize{
\begin{tabular}{rcrl}
Starting $j$ & Convergence to & Iterations &  \\
\hline
800 & 0.0001945544 & 23 & Down\\
500 & 0.0001795544 & 10 & Down\\
300 & 0.0001345544 &  5 & Down\\
200 & 0.0001195544 &  2 & Down\\
170 & 0.0001045544 &  2 & Down\\
160 & 0.0001045544 &  2 & Down\\
155 & 0.0001045544 &  2 & Up\\
152 & 0.0001045544 &  2 & Up\\
150 & 0.0001045544 &  2 & Up\\
\end{tabular}
}
\label{LN11 p=0.0001}
\end{table}
A sensible estimate of $p=0.0001$ is the average from the last 5 entries which gives 
$\hat{p}=0.0001045544$
with absolute error of $4.5544 \times 10^{-6}$.

\newpage
\subsubsection{Lognormal(0,1)}

\begin{table}[htbp]
\centering
\caption{\scriptsize
${\bf p=0.001}$, $\bfX_0 \sim LN(0,1)$, $\bfX_1 \sim Unif(0,50)$, $\max(\bfX_0)=11.86797$,
$T=21.98218$, $n_0=n_1=100$, $h=(x,\log x)$, $p$-increment $0.0001$.}
\scriptsize{
\begin{tabular}{rcrl}
Starting $j$ & Convergence to & Iterations &  \\
\hline
1000 & 0.001099445 & 19 & Down\\
 900 & 0.001099445 &  5 & Down\\
 820 & 0.001099445 &  2 & Down\\
 800 & 0.000999444 &  3 & Down\\
 790 & 0.000999444 &  2 & Down\\
 780 & 0.000999444 &  2 & Up\\
 770 & 0.000999444 &  2 & Up\\
 760 & 0.001099445 &  4 & Up\\ 
\end{tabular}
}
\label{LN01 p=0.001}
\end{table}
A sensible estimate of $p=0.001$ is the average from the last 5 entries which gives 
$\hat{p}=0.001019444$
with absolute error of $1.9444 \times 10^{-5}$.

\begin{table}[htbp]
\centering
\caption{\scriptsize
${\bf p=0.000}1$, $\bfX_0 \sim LN(0,1)$, $\bfX_1 \sim Unif(0,70)$, $\max(\bfX_0)=13.77121$,
$T=41.22383$, $n_0=n_1=100$, $h=(x,\log x)$, $p$-increment $0.000015$.}
\scriptsize{
\begin{tabular}{rcrl}
Starting $j$ & Convergence to & Iterations &  \\
\hline
900 & 0.0002392241 & 28 & Down\\
800 & 0.0001042241 & 25 & Down\\
700 & 0.0001042241 & 18 & Down\\
500 & 0.0001192241 &  6 & Down\\
360 & 0.0001042241 &  2 & Down\\
355 & 0.0001042241 &  2 & Up\\
350 & 0.0001042241 &  2 & Up\\
350 & 0.0001042241 &  2 & Up\\ 
\end{tabular}
}
\label{LN01 p=0.0001}
\end{table}
A sensible estimate of $p=0.0001$ is the average from the last 4 entries which gives 
$\hat{p}=0.0001042241$
with absolute error of $4.2241 \times 10^{-6}$.

\newpage
\subsubsection{f(2,7)}

\begin{table}[htbp]
\centering
\caption{\scriptsize
${\bf p=0.001}$, $\bfX_0 \sim f(2,7)$, $\bfX_1 \sim Unif(0,50)$, $\max(\bfX_0)=12.25072$,
$T=21.689$, $n_0=n_1=100$, $h=(x,\log x)$, $p$-increment $0.0001$.}
\scriptsize{
\begin{tabular}{rcrl}
Starting $j$ & Convergence to & Iterations &  \\
\hline
500 & 0.001103351 & 10 & Down\\
450 & 0.001003351 &  9 & Down\\
400 & 0.001003351 &  7 & Down\\
300 & 0.001003351 &  4 & Down\\
210 & 0.001003351 &  2 & Up\\
190 & 0.000903350 &  2 & Up\\
180 & 0.000903350 &  2 & Up\\
\end{tabular}
}
\label{f27 p=0.001}
\end{table}
A sensible estimate of $p=0.001$ occurs at the Down-Up shift which gives 
$\hat{p}=0.001003351$
with absolute error of $3.351 \times 10^{-6}$.

\begin{table}[htbp]
\centering
\caption{\scriptsize
${\bf p=0.0001}$, $\bfX_0 \sim f(2,7)$, $\bfX_1 \sim Unif(0,70)$, $\max(\bfX_0)=14.62357$,
$T=45.13234$, $n_0=n_1=100$, $h=(x,\log x)$, $p$-increment $0.000015$.}
\scriptsize{
\begin{tabular}{rcrl}
Starting $j$ & Convergence to & Iterations &  \\
\hline
750 & 0.0001341104  & 3 & Down\\
740 & 0.0001041104  & 5 & Down\\
730 & 0.0001041104  & 4 & Down\\
700 & 0.0001341104  & 3 & Up\\
660 & 0.0001041104  & 2 & Down\\
650 & 0.0001041104  & 2 & Up\\
645 & 0.0001041104  & 2 & Up\\
640 & 0.0001041104  & 3 & Up\\
\end{tabular}
}
\label{f27 p=0.0001}
\end{table}
A sensible estimate of $p=0.0001$ occurs at the Down-Up shift which gives
$\hat{p}=0.0001041104$
with absolute error of $4.1104 \times 10^{-6}$.

\newpage
\subsubsection{Weibull(0.8,2)}

\begin{table}[htbp]
\centering
\caption{\scriptsize
${\bf p=0.001}$, $\bfX_0 \sim Weibull(0.8,2)$, $\bfX_1 \sim Unif(0,40)$, $\max(\bfX_0)=8.081707$,
$T=22.39758$, $n_0=n_1=100$, $h=(x,\log x)$, $p$-increment $0.0001$.}
\scriptsize{
\begin{tabular}{rcrl}
Starting $j$ & Convergence to & Iterations &  \\
\hline
1000 & 0.001899263 & 3 & Down\\
1000 & 0.001099263 & 8 & Down\\
 950 & 0.000999262 & 2 & Immediate\\
 950 & 0.000999262 & 2 & Up\\
 940 & 0.001099263 & 4 & Up\\
 940 & 0.000999262 & 3 & Up\\
\end{tabular}
}
\label{Weibull.8_2 p=0.001}
\end{table}
In the 3rd entry there was an immediate convergence.
A sensible estimate of $p=0.001$ is the average from the last 5 entries which gives 
$\hat{p}=0.001039261$
with absolute error of $3.9261 \times 10^{-5}$.

\begin{table}[htbp]
\centering
\caption{\scriptsize
${\bf p=0.0001}$, $\bfX_0 \sim Weibull(0.8,2)$, $\bfX_1 \sim Unif(0,50)$, $\max(\bfX_0)=12.20032$,
$T=32.09036$, $n_0=n_1=100$, $h=(x,\log x)$, $p$-increment $0.000015$.}
\scriptsize{
\begin{tabular}{rcrl}
Starting $j$ & Convergence to & Iterations &  \\
\hline
700 & 0.0002096393 & 21 & Down\\
400 & 0.0001196393 & 11 & Down\\
300 & 0.0001946393 &  2 & Down\\
200 & 0.0001046393 &  5 & Down\\
130 & 0.0001046393 &  2 & Down\\
125 & 0.0001046393 &  2 & Up\\
120 & 0.0001046393 &  2 & Up\\
115 & 0.0001046393 &  2 & Up\\
\end{tabular}
}
\label{Weibull.8_2 p=0.0001}
\end{table}
A sensible estimate of $p=0.0001$ is the average from the last 5 entries which gives 
$\hat{p}=0.0001046393$
with absolute error of $4.6393 \times 10^{-6}$.

\subsubsection{2,4,6-trichlorophenol (ug/L)}

We use trichlorophenol  data from \mbox{\tt{https://wwwn.cdc.gov/nchs/nhanes}}\\
(dubbed urx3tb). 
There are 2604 observations of which 
the proportion exceeding $T=9.5$ is $p = 0.001152074$. Consider the 2604 observations as a
population and the problem is to estimate $p$ from a sample $\bfX_0$ 
of size $n_0=100$ where $\max(\bfX_0) < T$.

\begin{table}[htbp]
\centering
\caption{\scriptsize
${\bf p = 0.001152074}$, $\bfX_0$ a trichlorophenol sample. 
$\bfX_1 \sim Unif(0,30)$, $\max(\bfX_0)=3$,
$T=9.5$, $n_0=n_1=100$, $h=(x,\log x)$, $p$-increment $0.0001$.}
\scriptsize{
\begin{tabular}{rcrl}
Starting $j$ & Convergence to & Iterations &  \\
\hline
840  &  0.001099096  &   8    &  Down\\
800  &  0.000999095  &   7    &  Down\\
760  &  0.000999095  &   4    &  Down\\
755  &  0.001099096  &   2    &  Down\\
750  &  0.001099096  &   2    &  Up\\
740  &  0.000999095  &   2    &  Up\\
735  &  0.000999095  &   2    &  Up\\
732  &  0.001099096  &   4    &  Up\\
\end{tabular}
}
\label{URX3TB_trichlorophenol_I  p=0.001}
\end{table}
The 8 estimates in Table \ref{URX3TB_trichlorophenol_I  p=0.001} with $\max(\bfX_0)=3$
seem to be in a neighborhood of the true $p=0.001152074$. 
Their average is  $ 0.001049096 \approx p$ with standard deviation of $0.5345278 \times 10^{-5}$. 
Note that the shift from down to up occurs at $0.001099096 \approx p$.  The 3rd quartile from 
10,000 $B$'s is 0.001225, suggesting a first $j$ around 800, as well as a reasonable first guess for $p$.

Repeating  the iterations with different 10,000 fusions
and a different $\bfX_0$ with $\max(\bfX_0)=4.6$,
we see from Table \ref{URX3TB_trichlorophenol_II  p=0.001}
that the shift from down to up occurs at $0.00119882$ close to $p$
with absolute error $4.6746 \times 10^{-5}$.
%\newpage
In this case
the median $0.001091$ from 10,000 $B$'s provides an approximation to $p$ . In general,
however, the 3rd quartile
(here 0.003386) is a more prudent first guess.

\begin{table}[htbp]
\centering
\caption{\scriptsize
${\bf p = 0.001152074}$, $\bfX_0$ a trichlorophenol sample. 
$\bfX_1 \sim Unif(0,30)$, $\max(\bfX_0)=4.6$,
$T=9.5$, $n_0=n_1=100$, $h=(x,\log x)$, $p$-increment $0.0001$.}
\scriptsize{
\begin{tabular}{rcrl}
Starting $j$ & Convergence to & Iterations &  \\
\hline
800  & 0.00119882 & 16 & Down\\
700  & 0.00189882 &  9 & Down\\
600  & 0.00119882 &  5 & Down\\
590  & 0.00119882 &  6 & Down\\
530  & 0.00119882 &  2 & Up\\
520  & 0.00109882 &  2 & Up\\
515  & 0.00109882 &  2 & Up\\  
\end{tabular}
}
\label{URX3TB_trichlorophenol_II  p=0.001}
\end{table}

\subsubsection{Mercury (mg/kg)}
\label{Mercury}

The mercury data consists of 8,266 observations 
of which the proportion exceeding $T=22.41$
is $p=0.001088797$. 
The data source is 
NOAA's National Status and Trends Data\\
\mbox{\tt{https://products.coastalscience.noaa.gov/nsandt\_data/data.aspx}}.\\
The results of 10,000 fusions of a reference sample $\bfX_0$ with 
$\bfX_1 \sim Unif(0,50)$ samples are summarized in Table \ref{Mercury_I  p=0.001}.
The shift from down to up occurs at 0.000999351 not far from $p=0.001088797$ and 
the median  from 10,000  $B$'s is 0.001049, close to the true $p$. 

\begin{table}[htbp]
\centering
\caption{\scriptsize
${\bf p = 0.001088797}$, $\bfX_0$ a mercury sample. 
$\bfX_1 \sim Unif(0,50)$, $\max(\bfX_0)=7.99$,
$T=22.41$, $n_0=n_1=100$, $h=(x,\log x)$, $p$-increment $0.0001$.}
\scriptsize{
\begin{tabular}{rcrl}
Starting $j$ & Convergence to & Iterations &  \\
\hline
800  & 0.001099352 & 14 & Down\\
700  & 0.001199352 &  8 & Down\\
600  & 0.000999351 &  5 & Down\\
500  & 0.000999351 &  2 & Down\\
490  & 0.000999351 &  2 & Up\\
480  & 0.000999351 &  2 & Up\\
470  & 0.000999351 &  2 & Up\\  
\end{tabular}
}
\label{Mercury_I  p=0.001}
\end{table}

Repeating the iterations with a different reference sample $\bfX_0$ as well as different 
10,000 fusions with $\bfX_1 \sim Unif(0,50)$ samples, we see from Table \ref{Mercury_II  p=0.001} 
that the shift from down to up occurs at 
0.001099501 very close to the true $p = 0.001088797$. The median from 10,000 $B$'s is
$0.001704$ giving an idea as to the magnitude of the true $p$.

\begin{table}[htbp]
\centering
\caption{\scriptsize
${\bf p = 0.001088797}$, $\bfX_0$ a mercury sample. 
$\bfX_1 \sim Unif(0,50)$, $\max(\bfX_0)=11.9$,
$T=22.41$, $n_0=n_1=100$, $h=(x,\log x)$, $p$-increment $0.0001$.}
\scriptsize{
\begin{tabular}{rcrl}
Starting $j$ & Convergence to & Iterations &  \\
\hline
800  & 0.001199501 & 15 & Down\\
700  & 0.001199501 & 12 & Down\\
500  & 0.001199501 &  6 & Down\\
400  & 0.001099501 &  2 & Down\\
390  & 0.001099501 &  2 & Up\\
380  & 0.001099501 &  2 & Up\\
375  & 0.001199501 &  3 & Up\\
360  & 0.001099501 &  3 & Up\\ 
\end{tabular}
}
\label{Mercury_II  p=0.001}
\end{table}

\subsubsection{Can the Method Fail?}

Problems might occur when $\max(\bfX_0)$ is too small or too large relative to $T$. A 
relatively small $\max(\bfX_0)$ indicates that the observed data are just too small, a problem that 
could be ameliorated by increasing $\bfX_0$. A large $\max(\bfX_0)$ indicates that the 
$``\bullet"$ point is too close to the lower end of the B-curve.
In that case the iterative method could fail
to produce converging ``Up" sequences.
If the reference sample is sufficiently large, the removal of few large observations creates a new 
smaller $\max(\bfX_0)$ which could push the point
upward along  the B-curve thus producing converging ``Up" sequences.\\
\\
As an example, consider the case of 10,000 fusions of $\bfX_0 \sim Gamma(1,0.05)$ with 
 $\bfX_1 \sim Unif(0,160)$, where ${\bf p=0.001}$, 
$T=138.1551$, $n_0=n_1=100$, $\bfh=(x,\log x)$, $p$-increment $0.0001$. 
From Figure 3, 
with
$\max(\bfX_0)=122.1429$ the $``\bullet"$ point is at the bottom end of the B-curve slightly above
$(1,B_{(1)})=(1,0.0008912)$. 
By removing the largest 3 observations from $\bfX_0$, 
the $``\bullet"$ point moves upward along the B-curve, and 
the iterations with 
the new $\max(\bfX_0)=56.4284976$ and smaller sample sizes of 
$n_0=n_1=97$ gave Down-Up sequences which converged readily to
$\hat{p} =0.00117879$. This is close to what was obtained in Table \ref{Gamma_1_0.05 p=0.001}
with $\bfX_1 \sim Unif(0,170)$; here $\bfX_1 \sim Unif(0,160)$.

\begin{figure}[htbp]
\begin{center}
\includegraphics[height=6.2cm,angle=-90]
{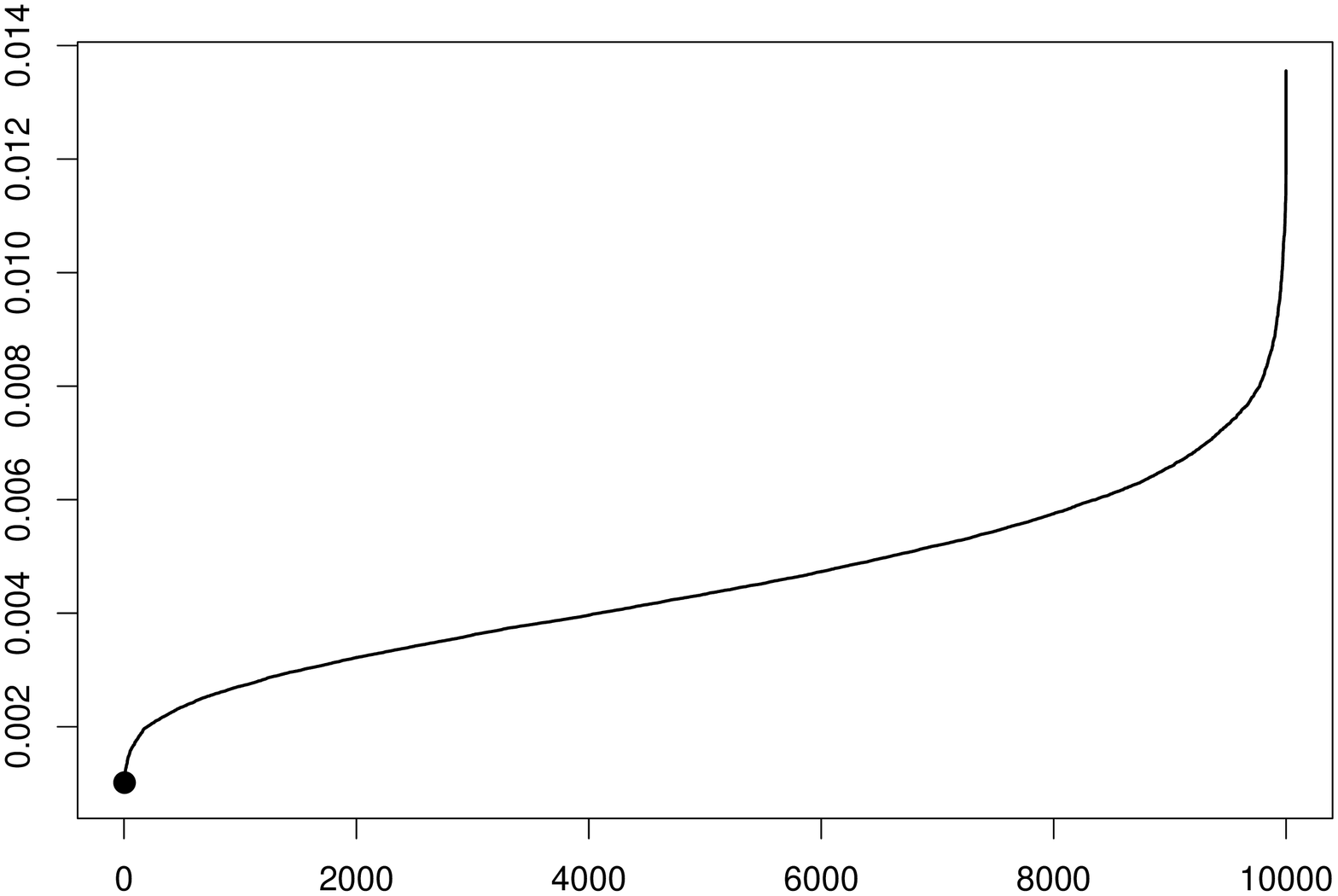}
\includegraphics[height=6.2cm,angle=-90]
{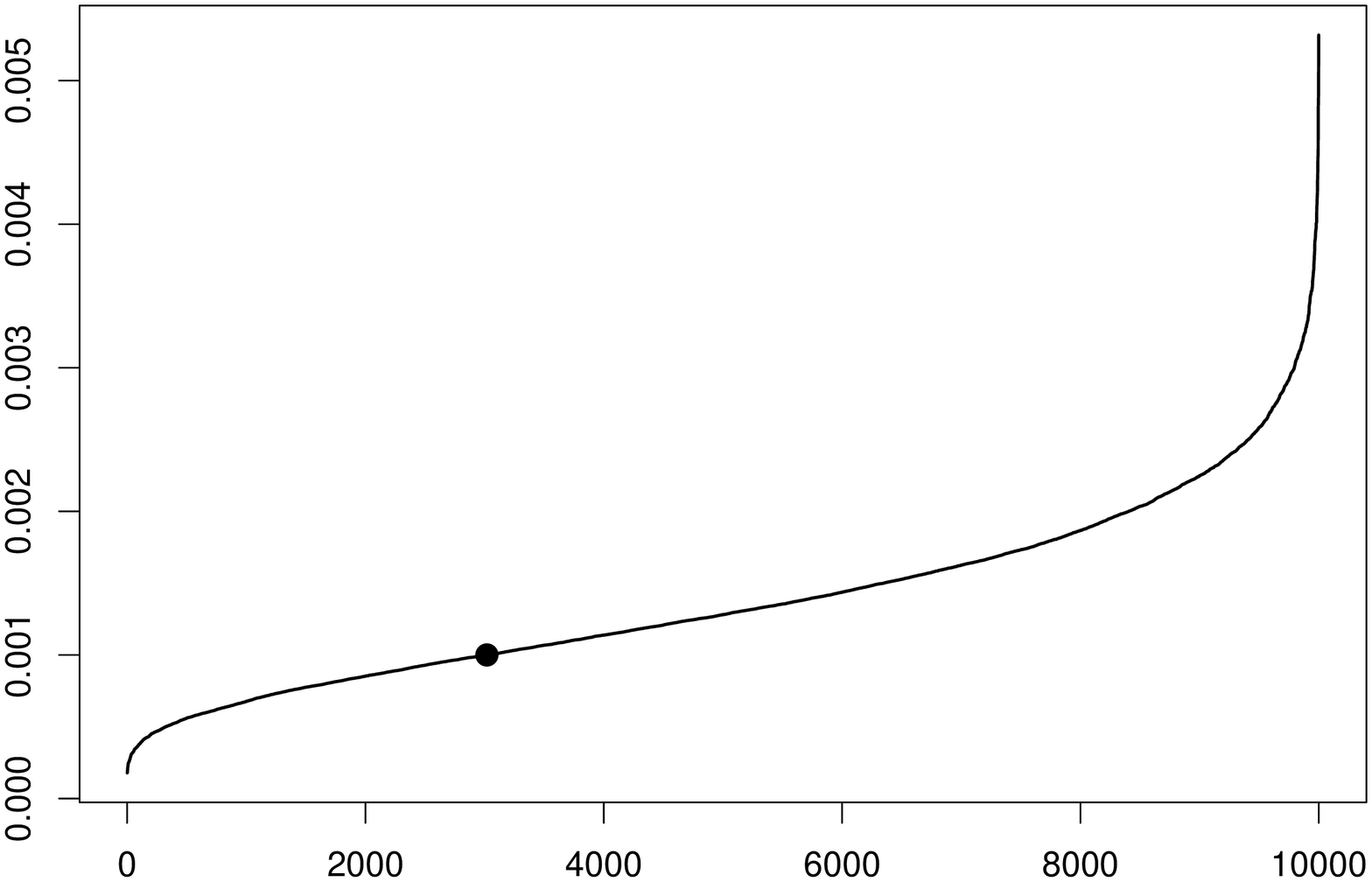}\\
\caption{B-Curves, 10,000 B's, from  Gamma(1,0.05), $p=0.001$, $T=138.1551$.
Left: 
$n_0=n_1=100$, $\max(\bfX_0)=122.1429$, the point ``$\bullet$" is at the bottom of the B-curve
slightly above $(1,B_{(1)})=(1,0.0008912)$.
Right: 
Three largest observations removed,
$n_0=n_1=97$, $\max(\bfX_0)=56.4284976$, the point ``$\bullet$" moves upward along the B-curve.}
%}
\end{center}
\label{B-Curves point moves upward}
\end{figure}

\section{Variability of Point Estimates}

Clearly, the iterative method (IM) can be repeated many times with 
different $B$-samples of size 1,000 taken from, say, 10,000 $B$'s (10,000 is our default) 
to produce tail probability estimates as above from which variance approximations can be
obtained. The following tables show typical variance approximations, obtained 
from  single convergent sequences where the starting 
$j$ corresponds to the 3rd quartile of the sampled 1,000 $B$'s. Each entry was obtained from 1,000
runs for both $n_0=n_1=100$ and $n_0=n_1=200$. There is improvement in precision going from
samples of size 100 to 200. In all cases reported here, and many other additional cases,
$\sigma_{\hat{p}}= \mbox{\cal{O}} (10^{-4})$. In the tables $\bar{p}$ is the average estimate of $p$
from 1,000 runs.

%%%%%%%%%%%%%%%%%%%%%%%%%%%%%%%%%%%%%%%%%%%%%%%%%%%%%%%%%%%%%%% Urx3tb

\begin{table}[htbt]
\centering
\caption{\scriptsize
$X_0 \sim \textbf{2,4,6-trichlorophenol (Urx3tb)}$: 
p = 1 - G(T) = 0.001152074, T = 9.50, $X_1 \sim $ Unif(0,12), $n_0 = n_1, h(x) = (x, \log x).$}
\scriptsize{
\begin{tabular}{|l|l|l|}
\hline
                             &  $\bar{p}$   & $\sigma_{\hat{p}}$   \\ \hline
$n_0 =100, \max(\bfX_0) = 8.8$  &  0.0011539   & 0.0004269           \\ \hline
$n_0 =200,\max(\bfX_0) =  8.8$  &  0.0011216   & 0.0002871            \\ \hline
\end{tabular}
}
\end{table}

%%%%%%%%%%%%%%%%%%%%%%%%%%%%%%%%%%%%%%%%%%%%%%%%%%%%%%%%%%%%%%% LN(0,1)

\vspace{-0.3in}
%\newpage
\begin{table}[htbp]
\centering
\caption{\scriptsize
$X_0 \sim \textbf{LN}(0,1)$ $: p = 1 - G(T) = 0.001, T = 21.98218, X_1 \sim $ Unif(1,60), 
$n_0 = n_1, h(x) = (x, \log x).$}
\scriptsize{
\begin{tabular}{|l|l|l|}
\hline
                                  &  $\bar{p}$   & $\sigma_{\hat{p}}$   \\ \hline
$n_0 =100, \max(\bfX_0) = 11.04102$  &  0.0011401   & 0.0004100            \\ \hline
$n_0 =200,\max(\bfX_0) =  11.04102$  &  0.0010598   & 0.0002823            \\ \hline
\end{tabular}
}
\end{table}

%%%%%%%%%%%%%%%%%%%%%%%%%%%%%%%%%%%%%%%%%%%%%%%%%%%%%%%%%%%%%%% Weibull

\begin{table}[htbp]
\centering
\caption{\scriptsize
$X_0 \sim \textbf{Weibull}(1,2)$ $: p = 1 - G(T) = 0.001, T = 13.81551,  X_1 \sim $ Unif(0,16), 
$n_0 = n_1, h(x) = (x, \log x).$}
\scriptsize{
\begin{tabular}{|l|l|l|}
\hline
                                     &  $\bar{p}$   & $\sigma_{\hat{p}}$   \\ \hline
 $n_0 =100, \max(\bfX_0) = 8.626444$    &  0.0011215   & 0.0001524            \\ \hline
 $n_0 =200,\max(\bfX_0) =  8.673713$    &  0.0010768   & 0.0001372             \\ \hline
\end{tabular}
}
\end{table}

%%%%%%%%%%%%%%%%%%%%%%%%%%%%%%%%%%%%%%%%%%%%%%%%%%%%%%%%%%%%%%% Pareto

\begin{table}[htbp]
\centering
\caption{\scriptsize
$X_0 \sim \textbf{Pareto}(1,4)$ $: p = 1 - G(T) = 0.001, T = 5.623413, X_1 \sim $ Unif(1,8), 
$n_0 = n_1, h(x) = (x, \log x).$}
\scriptsize{
\begin{tabular}{|l|l|l|}
\hline
                                   &  $\bar{p}$   & $\sigma_{\hat{p}}$   \\ \hline
$n_0 =100, \max(\bfX_0) = 3.08099$    &  0.0011549   & 0.0002256       \\ \hline
$n_0 =200,\max(\bfX_0) =  4.14516$    &  0.0009966   & 0.0002034         \\ \hline
\end{tabular}
}
\end{table}

\newpage
\section{Comparison: ROFS vs POT}

From a practical view point, some comparison is needed to assess the relative merit of ROSF/IM.
We provide in what follows a limited comparison against a well known method, however, a more extensive 
comparison is warranted and will be dealt with elsewhere.

Thus,
against the background provided in the previous sections, 
we compare two very different ways to obtain interval estimates for
small tail probabilities. The well known peaks over threshold 
(POT) based on extreme value theory, and the present iterative process 
based on repeated fusion of a given reference sample with external computer-generated 
uniformly distributed samples. 
The comparison is based on confidence interval coverage, interval width, and on the mean absolute error (MAE) which measures the discrepancy between $\hat{p}$ and the true tail probability $p$.
In Tables \ref{tb:T_1} to \ref{tb:Urx3tb}, $p$ is relatively small, $p=0.001$ (or approximately so), 
whereas in the last two Tables 
\ref{tb:F p=0.0001} and \ref{tb:LN(0,1) p=0.0001}, $p$ is smaller,
$p=0.0001$.
%\ref{tb:mercury p=0.0001},

Throughout the comparison the sample sizes are 
$n_0=n_1=100$ or $n_0=n_1=200$, and $\bfh(x)=(x,\log x)$. 
Thus, in the present comparison the reference $\bfX_0$ and the fusion samples $\bfX_1$
have size $n_0=100$ or $n_0=200$.

To save computation time, 
in each case of the iteration process
%$\hat{F}_B$ 
$F_B$ 
was obtained from 1000 fusions, and we use in each case a single convergent sequence where
the starting $j$ is such that 
%$B_{(j)} \approx \bar{B}$.
$B_{(j)}$ is approximately equal to the 3rd quartile of the observed 1000 $B$'s.
Starting at the 3rd quartile is computationally sensible as the corresponding $B_{(j)}$
most often converge to a point in a neighborhood of $p$ as $j$ increases. See Tables \ref{LN11 p=0.001}
to \ref{Mercury_II  p=0.001} and more examples in Kedem et al. (2018). 
%In most cases subsequent $B_{(j)}$ enter a
%neighborhood of $p$ and convergence occurs, as illustrated earlier.\\ 
%Starting too low might lead
%to convergence to a point lower than the true $p$.\\
%NOTE:
%STARTING WITH THE
%THIRD QUARTILE IS SENSIBLE AS IT IS USUALLY IN A  NEIGHBORHOOD SLIGHTLY ABOVE P.
%HENCE THERE IS A GOOD CHANCE OF ENTERING A NEIGHBORHOOD OF P AND ACHIEVE CONVERGENCE
%AS EXPLAINED ABOVE.
%IF WE START TOO LOW THEN WE COULD MISS P AND CONVERGE TO SOMETHING LESS THAN P.\\
\\
The following tables are the result of 500 runs. In each run the iteration method (IM) was 
repeated 500 times.

From the mean residual life (MRL) plots we obtained the thresholds $u$ needed for the POT method. In all cases reported in the tables, 
the MRL plots suggest the use of the largest $20\%$ of the reference data $\bfX_0$ for fitting the generalized Pareto (GP) distribution. 
We have noticed a deterioration in the POT results when using $30\%$, $15\%$ or $10\%$ of $\bfX_0$.
%As a check, similar results were obtained with 30\%!!!
The simulation details are given in 
Section \ref{Simulation Description} in the Appendix.\\
\\
An interesting picture emerges from Tables \ref{tb:T_1} to \ref{tb:LN(0,1) p=0.0001}.
%\ref{tb:Urx3tb}. 
For moderately large sample sizes of $n_0=100$ and $n_0=200$, 
regardless of the tail type, already with ${\bf N=50}$, that is,
the number of $\hat{p}$'s used in forming the CI 
for the true $p$ of the form $(\min(\hat{p}), \max(\hat{p}))$
(defined in Section \ref{Simulation Description}), 
the iteration process gives reliable and
relatively narrow confidence intervals, whereas the POT gives 
unacceptable coverage and in most cases wider CI's and greater MAE as well. 
The POT coverage increases significantly going from $n_0=100$ to
$n_0=200$, however, it seems that for the method to ``fire up" larger samples are needed.
Regarding ROSF,  the choice of ${\bf N=50}$ 
seems prudent across all cases,
and  with $n_0=200$
shorter CI's  achieve coverage similar to that from the smaller $n_0=100$. 
In all cases the MAE from the iteration process is much smaller than that obtained from POT.

\subsection{Comparison Tables}

The following tables compare ROSF and POT for $p = 0.001$ and $p = 0.0001$.

%%%%%%%%%%%%%%%%%%%%%%%%%%%%%%%%%%%%%%%%%%%%%%%%%%%%%%%%%%%%%%% T
\begin{table}[htbp]

\centering

\caption{$X_0 \sim \textbf{t}_{(1)} > 0 $ 
$: p = 1 - G(T) = 0.001, T = 631.8645, X_1 \sim $ Unif(0,800), $n_0 = n_1, 
\bfh(x) = (x, \log x).$ $p$-increment 0.0001.}

\scriptsize{

\begin{tabular}{ccccc|ccc}

                      &                         &\multicolumn{3}{c|}{$n_0 =100$}&\multicolumn{3}{c}{$n_0 =200$}                           \\ \cline{3-8} 

Method                &\multicolumn{1}{c|}{N}   &     Coverage         &      CI Length      &   MAE     &     Coverage   &      CI Length   &   MAE   \\ 
\hline

POT                   &\multicolumn{1}{c|}{-}   &     63.2\%           &      0.00372        &   0.00149 &    72.1\%      &    0.00292       &   0.00122\\ 
\hline              

% &\multicolumn{1}{c|}{5}   &     47.2\%           &      0.00098        &   0.00061 &    54.1\%             &    0.00079       &   0.00051\\ 

% &\multicolumn{1}{c|}{10}  &     57.2\%           &      0.00107        &   -       &    68.5\%             &    0.00093       &    -    \\

% &\multicolumn{1}{c|}{25}  &     74.3\%           &      0.00148        &   -       &    87.2\%             &    0.00125       &    -    \\

ROSF \& IM     &\multicolumn{1}{c|}{\bf 50}&     98.2\%           &      0.00213        &   0.00061 &    100\%       &    0.00193       &  0.00051   \\

               &\multicolumn{1}{c|}{100} &     100\%            &      0.00264        &   -       &    100\%          &    0.00241       &    -    \\

% &\multicolumn{1}{l|}{300} &     100\%            &      0.00321        &   -       &    100\%              &    0.00303       &    - 

\end{tabular}

}

\label{tb:T_1}

\end{table}

%%%%%%%%%%%%%%%%%%%%%%%%%%%%%%%%%%%%%%%%%%%%%%%%%%%%%%%%%%%%%%% Weibull
\begin{table}[htbp]
\centering
\caption{$X_0 \sim \textbf{Weibull}(1,2)$ $: p = 1 - G(T) = 0.001, T = 13.81551,  X_1 \sim $ Unif(0,16), $n_0 = n_1, \bfh(x) = (x, \log x).$ $p$-increment 0.00005.}
\scriptsize{
\begin{tabular}{ccccc|ccc}
                  &                         &\multicolumn{3}{c|}{$n_0 =100$}&\multicolumn{3}{c}{$n_0 =200$}                           \\ \cline{3-8} 
Method            &\multicolumn{1}{c|}{N}   &     Coverage         &      CI Length      &   MAE     &     Coverage          &      CI Length   &   MAE   \\ \hline
POT               &\multicolumn{1}{c|}{-}   &     82.7\%           &      0.00431        &   0.00131 &    87.8\%             &    0.00333       &   0.00083     \\ \hline
% &\multicolumn{1}{c|}{5}   &     43.2\%           &      0.00040        &   0.00068 &    52.4\%             &    0.00042       &   0.00051    \\ 
% &\multicolumn{1}{c|}{10}  &     65.2\%           &      0.00083        &   -       &    72.7\%             &    0.00091       &    -    \\
%&\multicolumn{1}{c|}{25}  &     84.2\%           &      0.00159        &   -       &    85.6\%             &    0.00154       &    -    \\
ROSF \& IM
             &\multicolumn{1}{c|}{\bf 50}  &     92.5\%             &      0.00287        & 0.00068       &    92.8\%       &    0.00231          &    -    \\
                  &\multicolumn{1}{c|}{100} &     100\%            &      0.00381        &   -       &    100\%              &    0.00321       &    -    \\
% &\multicolumn{1}{l|}{300} &     100\%            &      0.00506        &   -       &    100\%              &    0.00402       &    - 
\end{tabular}
}
\label{tb:Weibull_1_2}
\end{table}

%%%%%%%%%%%%%%%%%%%%%%%%%%%%%%%%%%%%%%%%%%%%%%%%%%%%%%%%%%%%%%% Pareto

\begin{table}[htbp]
\centering
\caption{$X_0 \sim \textbf{Pareto}(1,4)$ $: p = 1 - G(T) = 0.001, T = 5.623413, X_1 \sim $ Unif(1,8), $n_0 = n_1, \bfh(x) = (x, \log x).$ $p$-increment 0.0001.}
\scriptsize{
\begin{tabular}{ccccc|ccc}
                      &                         &\multicolumn{3}{c|}{$n_0 =100$}&\multicolumn{3}{c}{$n_0 =200$}                           \\ \cline{3-8} 
Method                &\multicolumn{1}{c|}{N}   &     Coverage         &      CI Length      &   MAE     &     Coverage          &      CI Length   &   MAE   \\ \hline
POT                   
&\multicolumn{1}{c|}{-}   &     81.8\%           &      0.00419        &   0.00121 &    84.5\%             &    0.00337       &   0.00070     
\\ \hline
% &\multicolumn{1}{c|}{5}   &     59.1\%           &      0.00068        &   0.00052 &    62.4\%             &    0.00066       &   0.00041      \ 
% &\multicolumn{1}{c|}{10}  &     66.7\%           &      0.00093        &   -       &    74.8\%             &    0.00091       &    -    \\
%&\multicolumn{1}{c|}{25}  &     84.1\%           &      0.00154        &   -       &    86.1\%             &    0.00148       &    -    \\
ROSF \& IM
&\multicolumn{1}{c|}{\bf 50}  &     96.2\%     &      0.00232        &   0.00052       &    97.8\%             &    0.00231       &   0.00041    \\
                      &\multicolumn{1}{c|}{100} &     100\%            &      0.00272        &   -       &    100\%              &    0.00269       &    -    \\
%&\multicolumn{1}{l|}{300} &     100\%            &      0.00397        &   -       &    100\%              &    0.00377       &    - 
\end{tabular}
}
\label{tb:Pareto_1_4}
\end{table}

%%%%%%%%%%%%%%%%%%%%%%%%%%%%%%%%%%%%%%%%%%%%%%%%%%%%%%%%%%%%%%% Gamma

\begin{table}[htbp]
\centering
\caption{$X_0 \sim \textbf{Gamma}(3,1)$ $: p = 1 - G(T) = 0.001, T = 11.22887, X_1 \sim $ Unif(0,20), $n_0 = n_1, \bfh(x) = (x, \log x).$ $p$-increment 0.00005.}
\scriptsize{
\begin{tabular}{ccccc|ccc}
                      &                         &\multicolumn{3}{c|}{$n_0 =100$}&\multicolumn{3}{c}{$n_0 =200$}                           \\ \cline{3-8} 
Method                &\multicolumn{1}{c|}{N}   &     Coverage         &      CI Length      &   MAE     &     Coverage          &      CI Length   &   MAE   \\ \hline
POT                   
&\multicolumn{1}{c|}{-}   &     77.3\%           &      0.00410        &   0.00096 &    86.1\%             &    0.00321       &   0.00081     \\ \hline              
%&\multicolumn{1}{c|}{5}   &     41.2\%           &      0.00057        &   0.00054 &    47.1\%             &    0.00056       &   0.00043      \\ 
%&\multicolumn{1}{c|}{10}  &     49.6\%           &      0.00093        &   -       &    56.6\%             &    0.00092       &    -    \\
%&\multicolumn{1}{c|}{25}  &     73.2\%           &      0.00137        &   -       &    82.8\%             &    0.00129       &    -    \\
ROSF \& IM
&\multicolumn{1}{c|}{\bf 50}  &     93.4\%     &      0.00188        &   0.00054       &    94.5\%             &    0.00175       &    0.00043    \\
                      &\multicolumn{1}{c|}{100} &     100\%            &      0.00256        &   -       &    100\%              &    0.00248       &    -    \\
%&\multicolumn{1}{l|}{300} &     100\%            &      0.00338        &   -       &    100\%              &    0.00315       &    - 
\end{tabular}
}
\label{tb:Gamma_3_1}
\end{table}

%%%%%%%%%%%%%%%%%%%%%%%%%%%%%%%%%%%%%%%%%%%%%%%%%%%%%%%%%%%%%%% IG(2,40)

\begin{table}[htbp]
\centering
\caption{$X_0 \sim \textbf{IG}(2,40)$ $: p = 1 - G(T) = 0.001, T = 3.835791, X_1 \sim $ Unif(0,8), $n_0 = n_1, \bfh(x) = (x, \log x).$ $p$-increment 0.00005.}
\scriptsize{
\begin{tabular}{ccccc|ccc}
                      &                         &\multicolumn{3}{c|}{$n_0 =100$}&\multicolumn{3}{c}{$n_0 =200$}                           \\ \cline{3-8} 
Method                &\multicolumn{1}{c|}{N}   &     Coverage         &      CI Length      &   MAE     &     Coverage          &      CI Length   &   MAE   \\ \hline
POT                   &\multicolumn{1}{c|}{-}   &     69.6\%           &      0.00324        &   0.00123 &    82.3\%             &    0.00316       &   0.00092     \\ \hline              
%&\multicolumn{1}{c|}{5}   &     59.4\%           &      0.00098        &   0.00047 &    85.6\%             &    0.00091       &   0.00041     \\ 
%&\multicolumn{1}{c|}{10}  &     76.9\%           &      0.00147        &   -       &    96.3\%             &    0.00133       &    -    \\
%&\multicolumn{1}{c|}{25}  &     89.9\%           &      0.00255        &   -       &    100\%              &    0.00147       &    -    \\
ROSF \& IM 
&\multicolumn{1}{c|}{\bf 50}  &     100\%      &      0.00289        &   0.00047       &    100\%              &    0.00206       &  0.00041    \\
                      &\multicolumn{1}{c|}{100} &     100\%            &      0.00332        &   -       &    100\%              &    0.00313       &    -    \\
%&\multicolumn{1}{l|}{300} &     100\%            &      0.00401        &   -       &    100\%              &    0.00371       &    - 
\end{tabular}
}
\label{tb:IG_2_40}
\end{table}

%%%%%%%%%%%%%%%%%%%%%%%%%%%%%%%%%%%%%%%%%%%%%%%%%%%%%%%%%%%%%%% LN(1,0)

\begin{table}[htbp]
\centering
\caption{$X_0 \sim \textbf{LN}(0,1)$ $: p = 1 - G(T) = 0.001, T = 21.98218, X_1 \sim $ Unif(1,60), $n_0 = n_1, \bfh(x) = (x, \log x).$ $p$-increment 0.00005.}
\scriptsize{
\begin{tabular}{ccccc|ccc}
                       &                         &\multicolumn{3}{c|}{$n_0 =100$}&\multicolumn{3}{c}{$n_0 =200$}                           \\ \cline{3-8} 
Method                 &\multicolumn{1}{c|}{N}   &     Coverage         &      CI Length      &   MAE     &     Coverage          &      CI Length   &   MAE   \\ \hline
POT                    &\multicolumn{1}{c|}{-}   &     81.5\%           &      0.00451        &   0.00111 &    85.2\%             &    0.00392       &   0.00103     \\ \hline              
%&\multicolumn{1}{c|}{5}   &     81.5\%           &      0.00121        &   0.00047 &    83.6\%             &    0.00108       &   0.00039      \\ 
%&\multicolumn{1}{c|}{10}  &     88.7\%           &      0.00169        &   -       &    90.4\%             &    0.00141       &    -    \\
%&\multicolumn{1}{c|}{25}  &     95.3\%           &      0.00191        &   -       &    98.1\%             &    0.00173       &    -    \\
ROSF \& IM 
                 &\multicolumn{1}{c|}{\bf 50}  &     100\%              &      0.00234        &   0.00047 &    100\%              &    0.00199       &  0.00039    \\
                       &\multicolumn{1}{c|}{100} &     100\%            &      0.00267        &   -       &    100\%              &    0.00244       &    -    \\
%&\multicolumn{1}{l|}{300} &     100\%            &      0.00301        &   -       &    100\%              &    0.00283       &    - 
\end{tabular}
}
\label{tb:LN_1_0}
\end{table}

%%%%%%%%%%%%%%%%%%%%%%%%%%%%%%%%%%%%%%%%%%%%%%%%%%%%%%%%%%%%%%% LN(1,1)

\begin{table}[htbp]
\centering
\caption{$X_0 \sim \textbf{LN}(1,1)$ $: p = 1 - G(T) = 0.001, T = 59.75377, X_1 \sim $ Unif(1,140), $n_0 = n_1, \bfh(x) = (x, \log x).$ $p$-increment 0.0001.}
\scriptsize{
\begin{tabular}{ccccc|ccc}
                       &                         &\multicolumn{3}{c|}{$n_0 =100$}&\multicolumn{3}{c}{$n_0 =200$}                           \\ \cline{3-8} 
Method                 &\multicolumn{1}{c|}{N}   &     Coverage         &      CI Length      &   MAE     &     Coverage          &      CI Length   &   MAE   \\ \hline
POT                    &\multicolumn{1}{c|}{-}   &     81.4\%           &      0.00435        &   0.00117 &    86.8\%             &    0.00399       &   0.00099     \\ \hline              
%&\multicolumn{1}{c|}{5}   &     43.7\%           &      0.00078        &   0.00069 &    53.2\%             &    0.00071       &   0.00052 \\ 
%&\multicolumn{1}{c|}{10}  &     56.9\%           &      0.00109        &   -       &    68.1\%             &    0.00099       &    -    \\
%&\multicolumn{1}{c|}{25}  &     79.6\%           &      0.00143        &   -       &    89.7\%             &    0.00121       &    -    \\
ROSF \& IM
                  &\multicolumn{1}{c|}{\bf 50}  &     89.1\%            &      0.00187        &   0.00069 &    100\%              &    0.00164       & 0.00052 \\
                       &\multicolumn{1}{c|}{100} &     100\%            &      0.00199        &   -       &    100\%              &    0.00192       &    -    \\
%&\multicolumn{1}{l|}{300} &     100\%            &      0.00243        &   -       &    100\%              &    0.00234       &    - 
\end{tabular}
}
\label{tb:LN_1_1}
\end{table}

%%%%%%%%%%%%%%%%%%%%%%%%%%%%%%%%%%%%%%%%%%%%%%%%%%%%%%%%%%%%%%% mercury

\begin{table}[htbp]
\centering
\caption{$X_0 \sim$ Mercury $: p = 1 - G(T) = 0.001088797, 
T = 22.41, X_1 \sim $ Unif(0,50), $n_0 = n_1, \bfh(x) = (x, \log x).$ $p$-increment 0.0001.}
\scriptsize{
\begin{tabular}{ccccc|ccc}
                       &                         &\multicolumn{3}{c|}{$n_0 =100$}&\multicolumn{3}{c}{$n_0 =200$}                           \\ \cline{3-8} 
Method                 &\multicolumn{1}{c|}{N}   &     Coverage         &      CI Length      &   MAE     &     Coverage          &      CI Length   &   MAE   \\ \hline
POT                    &\multicolumn{1}{c|}{-}   &     85.3\%           &      0.00455        &   0.00130 &    88.6\%             &    0.00398       &   0.00122     \\ \hline              
%&\multicolumn{1}{c|}{5}   &     54.5\%           &      0.00073        &   0.00048 &    49.9\%             &    0.00063       &   0.00045     \\ 
%&\multicolumn{1}{c|}{10}  &     66.7\%           &      0.00095        &   -       &    76.7\%             &    0.00096       &    -    \\
%&\multicolumn{1}{c|}{25}  &     84.9\%           &      0.00157        &   -       &    96.7\%             &    0.00145       &    -    \\
ROSF \& IM
                  &\multicolumn{1}{c|}{\bf 50}  &     97.5\%            &      0.00215        & 0.00048   &    100\%              &    0.00197       &0.00045 \\
                       &\multicolumn{1}{c|}{100} &     100\%            &      0.00259        &   -       &    100\%              &    0.00238       &    -    \\
%&\multicolumn{1}{l|}{300} &     100\%            &      0.00337        &   -       &    100\%              &    0.00313       &    - 
\end{tabular}
}
\label{tb:mercury}
\end{table}

%%%%%%%%%%%%%%%%%%%%%%%%%%%%%%%%%%%%%%%%%%%%%%%%%%%%%%%%%%%%%%% Urx3tb

\begin{table}[htbp]
\centering
\caption{$X_0 \sim$ URX3TB $: p = 1 - G(T) = 0.001152074, 
T = 9.50, X_1 \sim $ Unif(0,12), $n_0 = n_1, \bfh(x) = (x, \log x).$ $p$-increment 0.0001.
Data source for  URX3TB - 2,4,6-trichlorophenol (ug/L):
\tt{https://wwwn.cdc.gov/nchs/nhanes}}
\scriptsize{
\begin{tabular}{ccccc|ccc}
                       &                         &\multicolumn{3}{c|}{$n_0 =100$}&\multicolumn{3}{c}{$n_0 =200$}                           \\ \cline{3-8} 
Method                 &\multicolumn{1}{c|}{N}   &     Coverage         &      CI Length      &   MAE     &     Coverage          &      CI Length   &   MAE   \\ \hline
POT                    &\multicolumn{1}{c|}{-}   &     81.1\%           &      0.00433        &   0.00143 &    87.1\%             &    0.00376       &   0.00123 \\ \hline              
%&\multicolumn{1}{c|}{5}   &     38.9\%           &      0.00078        &   0.00055 &    42.6\%             &    0.00071       &   0.00044 \\ 
%&\multicolumn{1}{c|}{10}  &     54.3\%           &      0.00094        &   -       &    61.8\%             &    0.00092      &    -    \\
%&\multicolumn{1}{c|}{25}  &     72.1\%           &      0.00131        &   -       &    81.7\%             &    0.00125       &    -    \\
ROSF \& IM
                  &\multicolumn{1}{c|}{\bf 50}  &     89.1\%            &      0.00179        & 0.00055   &    96.9\%             &    0.00177       & 0.00044 \\
                       &\multicolumn{1}{c|}{100} &     100\%            &      0.00241        &   -       &    100\%              &    0.00235       &    -    \\
%&\multicolumn{1}{l|}{300} &     100\%            &      0.00264        &   -       &    100\%              &    0.00259       &    - 
\end{tabular}
}
\label{tb:Urx3tb}
\end{table}

%\paragraph{Comparison Tables for $ p = 0.0001$}

%%%%%%%%%%%%%%%%%%%%%%%%%%%%%%%%%%%%%%%%%%%%%%%%%%%%%%%%%%%%%%% F

\begin{table}[htbp]
\centering
\caption{$X_0 \sim \textbf{F}(2,12)$ $: p = 1 - G(T) = 0.0001, T = 21.84953, X_1 \sim $ Unif(0,25), $n_0 = n_1, \bfh(x) = (x, \log x).$ $p$-increment 0.00001.}
\scriptsize{
\begin{tabular}{ccccc|ccc}
                      &                         &\multicolumn{3}{c|}{$n_0 =100$}&\multicolumn{3}{c}{$n_0 =200$}                           \\ \cline{3-8} 
Method                &\multicolumn{1}{c|}{N}   &     Coverage         &      CI Length      &   MAE     &     Coverage          &      CI Length   &   MAE   \\ \hline
POT                   &\multicolumn{1}{c|}{-}   &     71.4\%           &      0.00062        &   0.00052 &    81.6\%             &    0.00053       &   0.000045     \\ \hline              
%&\multicolumn{1}{c|}{5}   &     45.2\%           &      0.00021        &   0.00022 &    49.1\%             &    0.00017       &   0.000019      \\ 
%&\multicolumn{1}{c|}{10}  &     67.2\%           &      0.00033        &   -       &    77.1\%             &    0.00026       &    -    \\
%&\multicolumn{1}{c|}{25}  &     88.5\%           &      0.00045        &   -       &    89.3\%             &    0.00037       &    -    \\
ROSF \& IM
                 &\multicolumn{1}{c|}{\bf 50}  &     95.2\%            &      0.00059        & 0.00022  &    96.3\%             &    0.00052       & 0.000019  \\
                      &\multicolumn{1}{c|}{100} &     100\%            &      0.00082        &   -       &    100\%              &    0.00069       &    -    \\
%&\multicolumn{1}{l|}{300} &     100\%            &      0.00105        &   -       &    100\%              &    0.00087       &    - 
\end{tabular}
}
\label{tb:F p=0.0001}
\end{table}

%%%%%%%%%%%%%%%%%%%%%%%%%%%%%%%%%%%%%%%%%%%%%%%%%%%%%%%%%%%%%%% LN(1,0)

\begin{table}[htbp]
\centering
\caption{$X_0 \sim \textbf{LN}(0,1)$ $: p = 1 - G(T) = 0.0001, T = 41.22383, X_1 \sim $ Unif(1,60), $n_0 = n_1, \bfh(x) = (x, \log x).$ $p$-increment 0.00001.}
\scriptsize{
\begin{tabular}{ccccc|ccc}
                      &                         &\multicolumn{3}{c|}{$n_0 =100$}&\multicolumn{3}{c}{$n_0 =200$}                           \\ \cline{3-8} 
Method                &\multicolumn{1}{c|}{N}   &     Coverage         &      CI Length      &   MAE     &     Coverage          &      CI Length   &   MAE   \\ \hline
POT                   &\multicolumn{1}{c|}{-}   &     72.1\%           &      0.00064        &   0.00045 &    82.6\%             &    0.00047       &   0.000039     \\ \hline              
%&\multicolumn{1}{c|}{5}   &     55.2\%           &      0.00021        &   0.00021 &    69.1\%             &    0.00017       &   0.000017      \\ 
%&\multicolumn{1}{c|}{10}  &     77.2\%           &      0.00033        &   -       &    89.1\%             &    0.00020       &    -    \\
%&\multicolumn{1}{c|}{25}  &     98.5\%           &      0.00041        &   -       &    99.3\%             &    0.00034       &    -    \\
ROSF \& IM
                 &\multicolumn{1}{c|}{\bf 50}  &     100\%            &      0.00066         &0.00021   &    100\%              &    0.00057       & 0.000017  \\
                      &\multicolumn{1}{c|}{100} &     100\%            &      0.00083        &   -       &    100\%              &    0.00079       &    -    \\
%&\multicolumn{1}{l|}{300} &     100\%            &      0.00113        &   -       &    100\%              &    0.00094       &    - 
\end{tabular}
}
\label{tb:LN(0,1) p=0.0001}
\end{table}

\newpage
\section{Discussion}

The numerous number of fusions of a given reference sample with computer generated
samples gives rise to different observables including the upper bounds for a tail
probability $p$ that were used in the paper. 
The upper bounds, obtained from the combined 
real and artificial data, were mostly much larger than $p$, some were less than $p$,
but some among the multitude
of upper bounds essentially coincided with $p$ and they were identified  
rather closely using an iterative procedure.

We have illustrated that, across a fairly wide range of distributional tail types,
repeated fusion of a reference sample with externally 
generated uniform random data allowed us to gain information  about the tail
behavior beyond the threshold using the notion of B-curves 
coupled with a well known formula from order statistics. In neighborhoods of the true $p$,
the consequential 
Down-Up sequences tended to transition or
shift at points close to $p$, providing surprisingly close estimates.
We have seen that with sample sizes 
on the order of 100
we can in many cases estimate tail probabilities on the order of 1/10,000. 
It seems that larger samples are needed
for much smaller tail probabilities, and that the method could fail when $\max(\bfX_0)$ is exceedingly small
or exceedingly large relative
to the threshold $T$. That is, when the ``$\bullet$" point on the B-curve
is very close to one of the two ends of the curve.

Throughout the paper the fusion samples were 
uniform samples whose support contained $T$. That is, the upper limit of the support
exceeded $T$. But other than this, no guide for choosing the upper limits was provided. 
Experience, however, shows that different upper limits give similar results.

The ideas presented in this paper can be extended in a number of
ways. For example, using ``fake" data from distributions other than uniform, and using
different fusion mechanisms together with appropriate inferential methods
other than the semiparametric method used in the paper. That is, explore different ways
of connecting $\bfX_0$ and $\bfX_1$, other than by means of their distributions as expressed by the 
density ratio model.
%Estimating $K$ in (\ref{CI}) is another possible extension.

Reliable estimation of tail probabilities is important in numerous fields from finance to
geophysics to meteorology to 
the design of ships and to optics; see
Pelinovsky et al. (2008), and Solli et al. (2007).

%See what they say

%\newpage
\appendix
\section{Appendix}

The appendix addresses the density ratio model (\ref{density-ratio}) for $m+1$ data sources
discussed briefly in Section \ref{Getting Upper Bounds by Data Fusion}.

\subsection{Asymptotic Distribution of $\hat{G}(x)$}

Define $\alpha_0 \equiv 0, \beta_0 \equiv 0$, 
$w_j(x) = \exp(\alpha_j + \beta_j'h(x))$, $\rho_i=n_i/n_0$,  $j = 1, \ldots, m$.

Maximum likelihood estimates for all the parameters and $G(x)$ can
be obtained by maximizing the empirical likelihood  over
the class of step cumulative distribution functions with jumps at
the observed values $t_1,\ldots,t_n$ (Owen 2001).
%\cite{Owen2001}. 
Let $p_i = dG(t_i)$ be
the mass at $t_i$, for $i=1,\ldots,n$. Then the empirical likelihood
becomes
\begin{eqnarray}
\mathcal{L}(\bftheta,G) = \prod_{i=1}^n p_i 
\prod_{j=1}^{n_1}\exp(\alpha_1+\beta_1'h(x_{1j}))\cdots
\prod_{j=1}^{n_m}\exp(\alpha_m+\beta_m'h(x_{mj})).
\label{likelihood}
\end{eqnarray}
Maximizing $\mathcal{L}(\bftheta,G)$ subject to the constraints
\begin{eqnarray}
\label{constrains}
\sum_{i=1}^np_i = 1, \; \sum_{i=1}^np_i [w_1(t_i)-1] = 0, \ldots, 
\sum_{i=1}^np_i [w_m(t_i)-1] = 0
\end{eqnarray}
we obtain the desired estimates. In particular,
\begin{eqnarray}
\label{G-hat}
\hat{G}(t) = \frac{1}{n_0} \cdot
\sum_{i=1}^n \frac{I(t_i \leq t)}{1 + \rho_1 \exp(\hat{\alpha}_1 + 
\hat{\beta}_1' h(t_i))
+ \cdots + \rho_m \exp(\hat{\alpha}_m + \hat{\beta}_m ' h(t_i))},
\end{eqnarray}
where $I(t_i\leq t)$ equals one for $t_i\leq t$ and is zero, otherwise.
Similarly, $\hat{G}_j$ is estimated by summing 
$\exp(\hat{\alpha}_j + \hat{\beta}_j ' h(t_i))dG(t_i)$.

The asymptotic properties of the estimators have been studied by a number of authors
including Qin and Zhang (1997), Lu (2007), and Zhang (2000).
%\cite{QZ1997},\cite{Lu2007},\cite{Zhang2000}. 

Define the following quantities: $\bfrho = \mbox{diag}\{\rho_1, \ldots, \rho_m\}$, 
$$A_j(t) = \int \frac{w_j(y)I(y \leq t)}{\sum_{k=0}^m \rho_k w_k(y)} dG(y), \quad
B_j(t) = \int \frac{w_j(y) h(y)I(y \leq t)}{\sum_{k=0}^m \rho_k w_k(y)} dG(y),$$
$$
\bar{A}(t) = (A_1(t), \ldots, A_m(t))', \quad
\bar{B}(t) = (B'_1(t), \ldots, B'_m(t))'.
$$
Then the asymptotic distribution of $\hat{G}(t)$ for $m \geq 1$ is given by the following
result due to Lu (2007).
%\cite{Lu2007}.
{\Theorem 
%(Lu 2007.)
\label{Theorem-weak-convergence-G-hat-G} 
Assume that the sample size ratios
$\rho_j = n_j/n_0$ are positive and finite and remain fixed as the total sample size
$n = \sum_{j=0}^m n_j \rightarrow \infty$.
The
process $\sqrt{n}(\hat{G}(t)- G(t))$ converges to a zero-mean
Gaussian process in the space of real right continuous functions
that have left limits with covariance matrix given by}

%(\ref{cov-G-hat-G})

\begin{eqnarray}
\label{cov-G-hat-G}
\lefteqn{
\mbox{Cov} \{ \sqrt{n}(\hat{G}(t)- G(t)), \sqrt{n}(\hat{G}(s)- G(s)) \} =   }
\nonumber\\
& &
\left(\sum_{k=0}^m \rho_k \right)
\biggl(
G(t\wedge s) -G(t)G(s)
-  \sum_{j=1}^m \rho_j A_j(t\wedge s)
\biggr)       \nonumber\\
& &
+ \biggl(\bar{A}'(s)\bfrho, \bar{B}'(s)(\bfrho \otimes I_p) \biggr) S^{-1}
\left( \begin{array}{c}
\bfrho \bar{A}(t)  \\
(\bfrho \otimes I_p) \bar{B}(t)
\end{array} \right).
\end{eqnarray}
\noindent
{\em where
$I_p$ is the $p \times p$ identity matrix, and $\otimes$ denotes
Kronecker product.}\\ 
\\
For a complete proof see Lu (2007).
%\cite{Lu2007}. 
The proof for $m=1$ is given in Zhang (2000).
%\cite{Zhang2000}.

%From Theorem \ref{Theorem-weak-convergence-G-hat-G} we get pointwise  symmetric
%confidence intervals for $G(t)$ for each given $t$ using $\hat{G}(t)$. 
%Pointwise symmetric confidence intervals can be calculated by estimating the 
%covariance
%matrix (\ref{cov-G-hat-G}). Let 
Denote by $\hat{V}(t)$ the estimated variance of $\hat{G}(t)$ as given in
(\ref{cov-G-hat-G}). Replacing parameters by their estimates, 
a $1-\alpha$ level pointwise confidence interval 
for
$G(t)$ is approximated by
\begin{eqnarray}
\label{SP Lu CI-pointwise}
\left(
\hat{G}(t) - z_{\alpha/2}\sqrt{\hat{V}(t)}, \;\; \hat{G}(t) + z_{\alpha/2}
\sqrt{\hat{V}(t)}
\right),
\end{eqnarray}
where $z_{\alpha/2}$ is the upper $\alpha/2$ point of the standard normal distribution.
Hence, a $1-\alpha$ level pointwise confidence interval 
for $1-G(T)$ for any $T$, and in particular
for relatively large thresholds $T$ is approximated by 
\begin{eqnarray}
\bigg(1-\hat{G}(t)-z_{\alpha/2}\sqrt{\hat{V}(t)},
1-\hat{G}(t)+z_{\alpha/2}\sqrt{\hat{V}(t)}\bigg).
\label{SP Lu CI-pointwise for 1-G(T)}
\end{eqnarray}

\subsection{Simulation Description}
\label{Simulation Description}

The following steps were followed. There were 500 runs. 
In each run the iteration method (IM) was repeated 500 times.\\
\\
First, a reference $\bfX_0$ was obtained.\\ 
\\
POT:\\ 
The POT procedure was applied to get both an estimate $\hat{p}$ and a confidence interval (CI).
The MRL plots suggest the use of the largest $20\%$ of the reference data $\bfX_0$ for fitting the generalized Pareto (GP) distribution.\\ 
\\
ROSF/IM:\\
$\bfX_0$ was fused with $\bfX_1$ 1000 times (ROSF) to get $F_B$ and then $\hat{p}$ (IM).\\
$\bfX_0$ was fused again with different $\bfX_1$ 1000 times to get $F_B$ and $\hat{p}$.\\
%Fuse $\bfX_0$ again with different $\bfX_1$ 1000 times to get $F_B$ and $\hat{p}$.\\
This was repeated 500 times.\\ 
\\
The iterative method thus gave 500 $\hat{p}$'s. 
We then chose at random $N$ $\hat{p}$'s from 500 $\hat{p}$'s
to construct a CI for the true $p$ as $(\min(\hat{p}), \max(\hat{p}))$.\\
This is run 1. \\
\\ 
The above steps were repeated, for both POT and ROSF/IM each time with a different $\bfX_0$, 500 times 
(runs) to obtain coverage and average CI length. 
In the tables, CI length is an average length from 500 intervals.\\
\\
Since there are 500 runs, POT gave 500 $\hat{p}$'s.
Regarding IM, a single $\hat{p}$ was chosen at random (out of 500 $\hat{p}$'s)
from each of the 500 runs. 
The mean absolute error (MAE) was obtained in both cases  
from the mean of 500 absolute differences 
$\sum( |\hat{p_i} - p|)/500$, where $p=0.001$ or $p=0.0001$.
In the iterative method, in each table the MAE is reported 
once on the line corresponding to $N=50$.

%For all N = 5, 50, 100 etc. I choose hat{p} at random from 500 estimates.   F
%or each time,  I generate 500 estimates by our method. 1 estimates by POT. POT give us a 
%CI by formula.  for N = 5 , I sample 5 estimates from 500 estimates to get CI, 
%for N = 50, I sample 50 estimates from the same 500 estimates to get CI. etc. 
%I repeat this 500 times to compute the coverage and  MAE. 

\end{document}